\documentclass[twocolumn]{aastex63}
\usepackage{amsmath} 
\usepackage{graphicx} 
\usepackage{textcomp}
\usepackage{gensymb}
\usepackage{natbib} 
\usepackage{longtable}
\usepackage{multirow}
\usepackage{float}
\usepackage{enumitem}
\usepackage{rotating}

\begin{document}
\def\teff{$T_{eff}$ }
\def\av{$A_{V}$}
\graphicspath{{figures/}}

\turnoffedit
\received{}
\revised{}
\accepted{17 February 2022}
\submitjournal{ApJ}
\shorttitle{NIR Excess and Optical Veiling}
\shortauthors{Sullivan \& Kraus}

\title{Optical and Near-Infrared Excesses are Correlated in T Tauri Stars}

\author[0000-0001-6873-8501]{Kendall Sullivan}
\altaffiliation{NSF Graduate Research Fellow}
\affil{University of Texas at Austin, Austin TX 78712, USA}

\author{Adam L. Kraus}
\affil{University of Texas at Austin, Austin TX 78712, USA}

\correspondingauthor{Kendall Sullivan}
\email{kendallsullivan@utexas.edu}

\begin{abstract}
Accretion is one of the defining characteristics of classical T Tauri stars, fueled by the presence of a circumstellar disk comprised of dust and gas. Accretion produces a UV and optical excess, while re-radiated emission at the inner edge of the dust component of the disk produces a near-infrared (NIR) excess. The interplay between stars and their disks helps regulate protoplanetary disk evolution and dispersal, which is key to a full understanding of planet formation. To investigate the relations between NIR excess and optical excess in both single and binary stars, we used an archival sample of spectroscopically characterized members of the Taurus star-forming region ($\tau \sim$ 1-2 Myr) with measured luminosities, spectral types, and optical veiling. We combined the archival sample with 2MASS and WISE NIR photometry and high-resolution imaging surveys. We found that NIR and optical excesses are correlated in multiple NIR photometric bands, suggesting that they are closely related, likely because more massive disks have higher inner dust disk walls and are also associated with higher accretion rates. We also found that multiplicity has no impact on accretion or inner disk properties in a sample with a wide range of separations, but the sample was too small to specifically investigate close binaries, where the effects of multiplicity on disk properties should be most significant. 
\end{abstract}

\keywords{}

\section{Introduction} \label{sec:intro}
Classical T Tauri stars \citep[CTTS;][]{Joy1945} represent a key stage of star and planet formation, where the stellar photosphere is newly visible but the star retains its circumstellar disk, a remnant of its formation within the surrounding molecular cloud. The star and disk interact as the star irradiates the disk and the disk accretes onto the star, causing complex co-dependent star and disk evolution over 5-10 Myr (for single and wide binary stars: close binary disk lifetimes are closer to 1 Myr; \citealt{Haisch2001, Armitage2003, Cieza2007, Cieza2009, Kraus2012a}) as the disk accretes onto the central star, assembles planets, and photoevaporates.

Circumstellar disks have dust and gas components; the gas disk extends inward to the magnetospheric truncation radius, which occurs within a few R$_{\star}$ of the photosphere \citep{Koenigl1991, Akeson2005, Johnstone2014}. The gaseous component of circumstellar disks is thought to fall onto the surface of the host star via magnetospheric accretion, where gas from the disk funnels onto the stellar surface along magnetic field lines \citep[e.g.,][]{Uchida1985, Bertout1988, Koenigl1991, Calvet1998}. The process of gas accretion onto the star produces continuum emission in excess of the stellar radiation in the optical and ultraviolet \citep[UV; e.g.,][]{Basri1990, Hartigan1991, Calvet1998}, which is called veiling in reference to its apparent reduction of the depth of absorption lines in a continuum-normalized spectrum.

The dust disk is inwardly truncated around $R \sim 0.1$ au \citep{Muzerolle2003, Eisner2005, Eisner2007, Dullemond2010, Davies2020} because of dust sublimation, which is thought to occur at temperatures around 1,500-2,000 K \citep{Pollack1994} and results in a raised rim of dust at the dust sublimation radius that produces a large fraction of the infrared (IR) flux \citep{Folha2001, Johns-Krull2001, Muzerolle2003, Eisner2005, Eisner2007, Dullemond2010, Davies2020}. The absorption and re-radiation of incident stellar flux at the inner dust disk edge produces an IR continuum excess.

Clarity regarding the dynamics, time scale, and detailed physics of dust evolution in circumstellar disks is vital for an understanding of planetesimal and planet formation, which occurs from the dust in the disk. Over time, dust and planetesimals settle into the midplane of the disk and radially migrate inward toward the star \citep[e.g.,][]{Artymowicz1993, Korycansky1993, Ward1997}, partially setting a timescale for dust dissipation, while the dust sublimation radius sets the minimum radius for \textit{in situ} rocky planet formation. Features such as the mass accretion rate onto the star dictate the total luminosity irradiating the inner disk edge, and therefore impact disk evolution by controlling the location of the dust sublimation radius. Thus, understanding gas accretion, disk properties, and the relationship between those attributes of a system is vital for a clear understanding of the planet formation process.

Because the accretion process impacts both the host star and the circumstellar disk, the optical and IR excesses should be correlated. This hypothesis regarding the link between optical and NIR excess was tested by \citet{Hartigan1991}, who found that K-L and K-N IR excesses were both correlated with optical veiling, supporting the inner disk structure model and the relationship between disk excess in the IR and accretion continuum excess in the optical. Other work pushed blueward in the IR, finding that there was excess at the IYJH bands \citep{Cieza2005, Fischer2011} which originated from an unknown source but had a spectrum that was consistent with a warm-hot blackbody ($\sim$2200-5000K; \citealt{Fischer2011}). However, \citet{McClure2013} investigated the near-infrared (NIR) excess and found that their spectral energy distribution (SED) fits did not require the presence of the warm-hot blackbody component, instead finding that a two- to three-component fit consisting of blackbodies at hot (gas emission; $\sim$8000K), warm (dust sublimation; $\sim$1600K) and possibly cool (dust emission; $\sim$800K) temperatures was sufficient to describe the NIR excess emission. 

Thus, although the relationship between K band excess and optical veiling is relatively well-established, and it is known that J and H excesses are correlated with K excess \citep[e.g.,][]{Cieza2005}, the relationship between optical veiling (as a proxy for gas accretion) and the NIR SED (as a proxy for the radius and structure of the inner dust disk wall) is still not well understood. This is partially because past studies of these relations in the NIR have either been large heterogeneously observed and characterized samples with less detailed analysis, or labor-intensive in-depth analyses of a small number of systems. Additionally, well-calibrated all-sky NIR photometry is only recently available via WISE \citep{Wright2010}, and establishing the photospheric SEDs of disk-hosting stars (to measure the optical and NIR excesses) has required recent advances in our knowledge of young stars' photospheric emission \citep[e.g.,][]{Herczeg2014}. 

Another potential complicating factor in past analyses of disk and accretion properties in CTTS is multiplicity. Stellar multiplicity is ubiquitous: $\sim$50\% of Sun-like stars in the field are binaries \citep{Duquennoy1991, Raghavan2010}, with multiplicity fractions decreasing to $\sim$25\% for M stars in the field \citep[e.g.,][]{Winters2019}. Multiplicity rates for young star-forming regions are typically close to double those of the field \citep[e.g.,][]{Duchene2013}. Multiplicity impacts disk formation and evolution by decreasing the timescale for disk dissipation \citep[e.g.,][]{Cieza2009, Kraus2012a} and reducing disk masses \citep[e.g.,][]{Jensen1996, Harris2012, Zurlo2021}, which could affect disk emission and may impact accretion rate, which is correlated with disk mass \citep{Manara2016m}. Past works have typically attempted to screen out binaries, but undetected binaries will still bias measurements of properties such as stellar age, mass, and spectral type \citep[e.g.,][]{Furlan2020, Sullivan2021}. Beyond simple removal from samples because of potential biases, information about the disk emission from known binaries may inform our understanding of circumstellar disk structure and planet formation in binary star systems. Thus, it is important to not only identify binaries to facilitate their removal from samples of single stars, but also to study the properties of the binary star disk host population in their own right.

To study the relationships between NIR excesses and optical veiling in both single and binary stars, we have used a sample of $\sim$ 160 PMS stars in Taurus with homogeneously measured properties from \citealt{Herczeg2014} (hereafter HH14), including bolometric luminosity, extinction, spectral type, age, and optical veiling ($r_{7510}$), and a young age with only moderate dispersion ($\tau \sim 1-2$ Myr; e.g., \citealt{Krolikowski2021}). We have cross-matched the HH14 Taurus sample with the 2MASS \citep{Skrutskie2006} and ALLWISE \citep{Wright2010} surveys to construct near- to mid-IR ($1.2< \lambda < 22 \mu$m) SEDs and determine the disk-host status of all the stars in our sample. We have also cross-matched our sample with several high-contrast imaging surveys of Taurus members to identify any binaries in the HH14 sample that had not been identified at the time of that analysis. 

Using this large homogeneously-characterized sample, we have explored relationships between optical veiling and NIR excess, constructed disk SEDs for our sample, and examined differences in disk and accretion properties between single and binary stars. Section \ref{sec:sample} discusses our sample selection, and Section \ref{sec:excess} describes our calculation of the NIR excess. Section \ref{sec:veiling correlations} presents the relationships we found between optical veiling and NIR excess, Section \ref{sec:disc} discusses our results, and Section \ref{sec:conclusion} presents our conclusions.

\section{Sample Selection}\label{sec:sample}


\subsection{Initial Sample Selection}
We began with the TTS sample from HH14, which is a relatively homogeneous collection of moderate-resolution spectroscopic observations of 281 stars from the Taurus ($\tau \sim 1$ Myr; e.g., \citealt{Krolikowski2021}), Lupus ($\tau \sim 1-3$ Myr; \citealt{Galli2020}), Ophiuchus ($\tau \sim 2-6$ Myr; \citealt{Esplin2020}), TW Hydra ($\tau \sim 8-10$ Myr; \citealt{Venuti2019}), and MBM 12 ($\tau \sim 2$ Myr; \citealt{Luhman2001}) associations. To reduce the age spread in the population used for our analysis, we restricted the sample to Taurus targets. To ensure high-quality photometry, we removed known binaries (as marked in the HH14 target table) that had separate spectral observations but were unresolved in 2MASS, but retained systems that were either resolved in both photometry and spectroscopy or were unresolved in both.

The observations and data reduction for the selected Taurus sample are described in detail in HH14, but are briefly described here for completeness. HH14 obtained low-resolution spectra on the Hale 200-inch telescope at Palomar Observatory using the Double Spectrograph \citep[DBSP;][]{Oke1982} and on the Keck I telescope at Keck Observatory using the Low-Resolution Imaging Spectrograph \citep[LRIS;][]{Oke1995, McCarthy1998} between November 2006 and December 2008. The spectra were reduced and flux calibrated using spectrophotometric standards. Spectra of known binary stars with separations of $\rho < 5\arcsec$ were extracted simultaneously using PSF fitting of a reference source to extract one component's flux and then the other.

HH14 calculated the mass, spectral type, bolometric luminosity L$_{bol}$, and optical excess (veiling) at 7510 \AA\ (denoted as $r_{7510}$), among other properties of their sample. For clarity, we note here that $r_{\lambda} = \frac{F_{acc}}{F_{phot}}$ at a given wavelength $\lambda$, and can be measured as a fractional decrease in the depth of spectral lines relative to the normalized continuum, which is equivalent to comparing an observed, flux-calibrated spectrum to combined photospheric and accretion spectrum models. 

To construct our preliminary sample we selected all HH14 targets that fit the above criteria and were not marked as ``continuum'' objects (objects without visible spectral lines). We were left with 175 targets. If an object had multiple veiling measurements we took the median of the values. HH14 found that the veiling did not change by more than a factor of three for systems with multiple veiling measurements, and that strongly veiled systems never became weakly veiled (or vice versa). We assume that the median is roughly representative of the veiling value, but note that there may be some scatter in the veiling-NIR excess relation because of variable veiling.

HH14 performed their calculations based on an assumed distance to each system, typically derived using a parallax for a nearby star. The majority of our targets were included in the \textit{Gaia} EDR3 catalog \citep{Gaia2016, Gaia2021}, so when possible we used the distances calculated by inverting the \textit{Gaia} parallax. 14 systems did not have a \textit{Gaia} match, and for those systems we used the HH14 distance. In HH14 the assumed distance for each system was dependent on its location in Taurus: the assumed distance was 131 pc for stars near the Lynds 1495 complex \citep{Torres2012}; 147 pc for stars near T Tau \citep{Loinard2007}; 161 pc for stars near HP Tau \citep{Torres2009}; and 140 pc for all other Taurus objects. Because the calculation of NIR excesses was dependent on the system luminosity, which was calculated by HH14 using their original distances, we rescaled the luminosity to the correct distance for systems with \textit{Gaia} parallaxes.

The names, 2MASS source names, veiling values, distances, and \av\ values for our targets are listed in Table \ref{tab:source_params}.

\subsection{Identifying NIR Photometry}
\begin{figure*}
\gridline{\fig{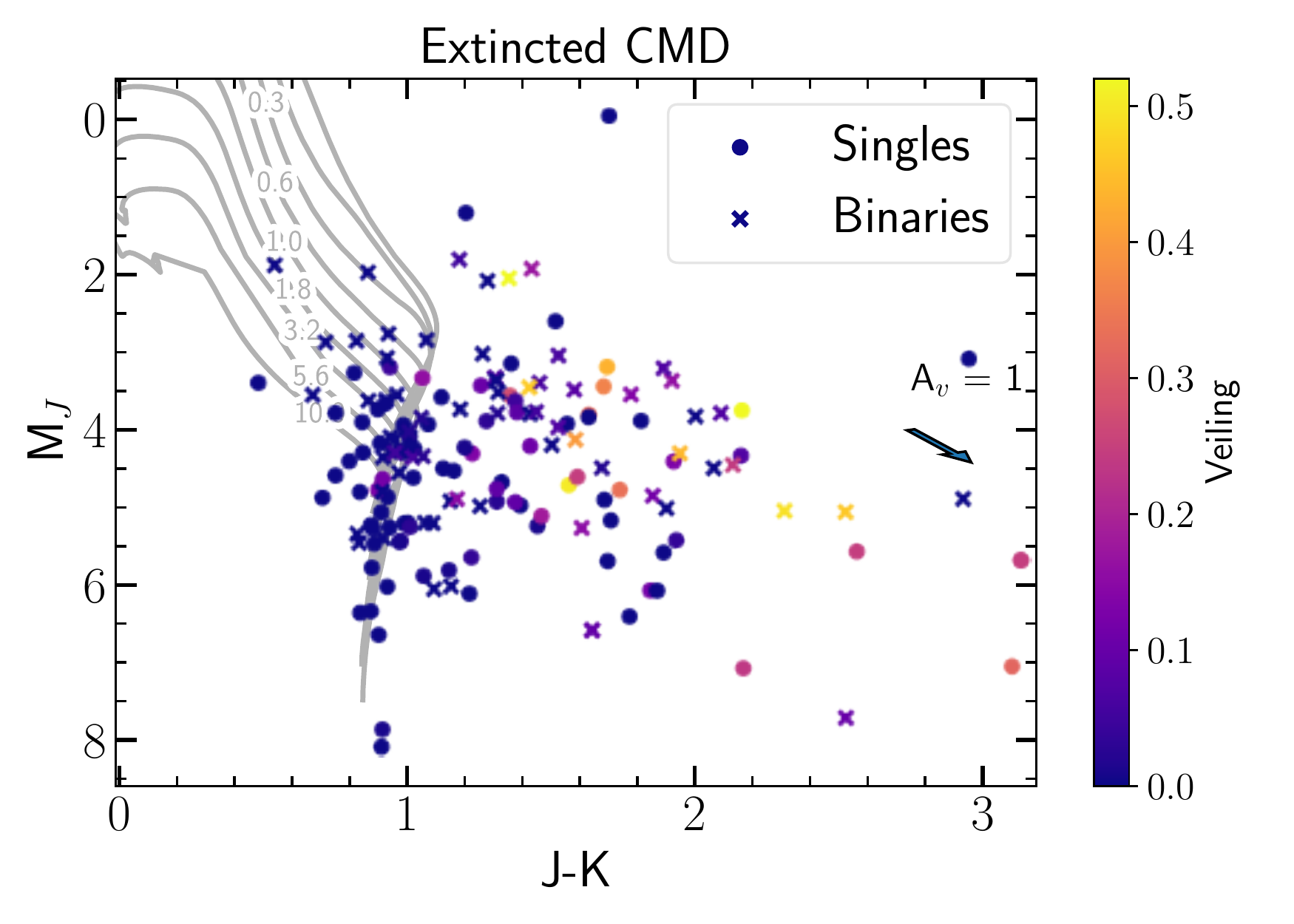}{0.45\linewidth}{}
    \fig{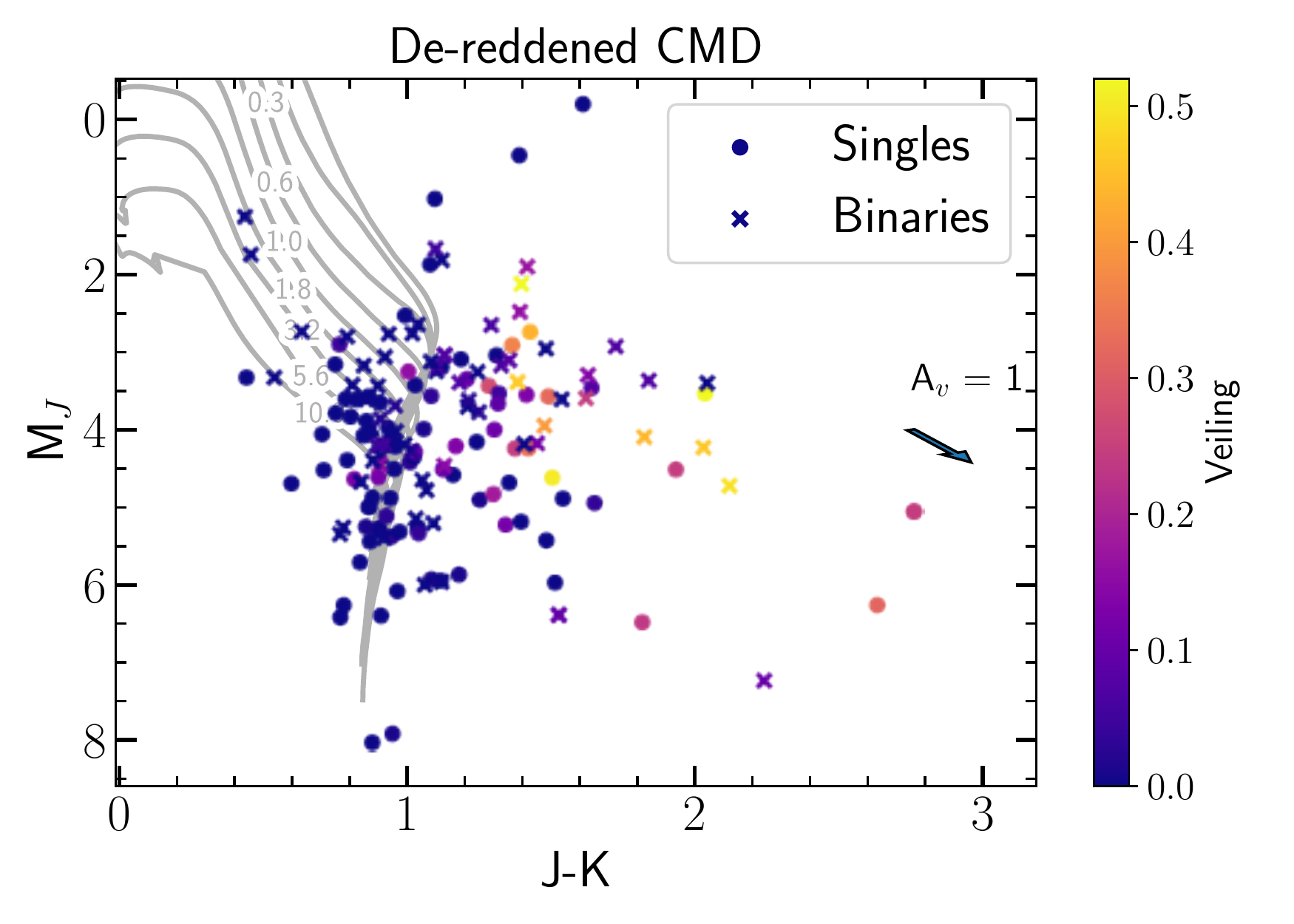}{0.45\linewidth}{}
	}
\gridline{\fig{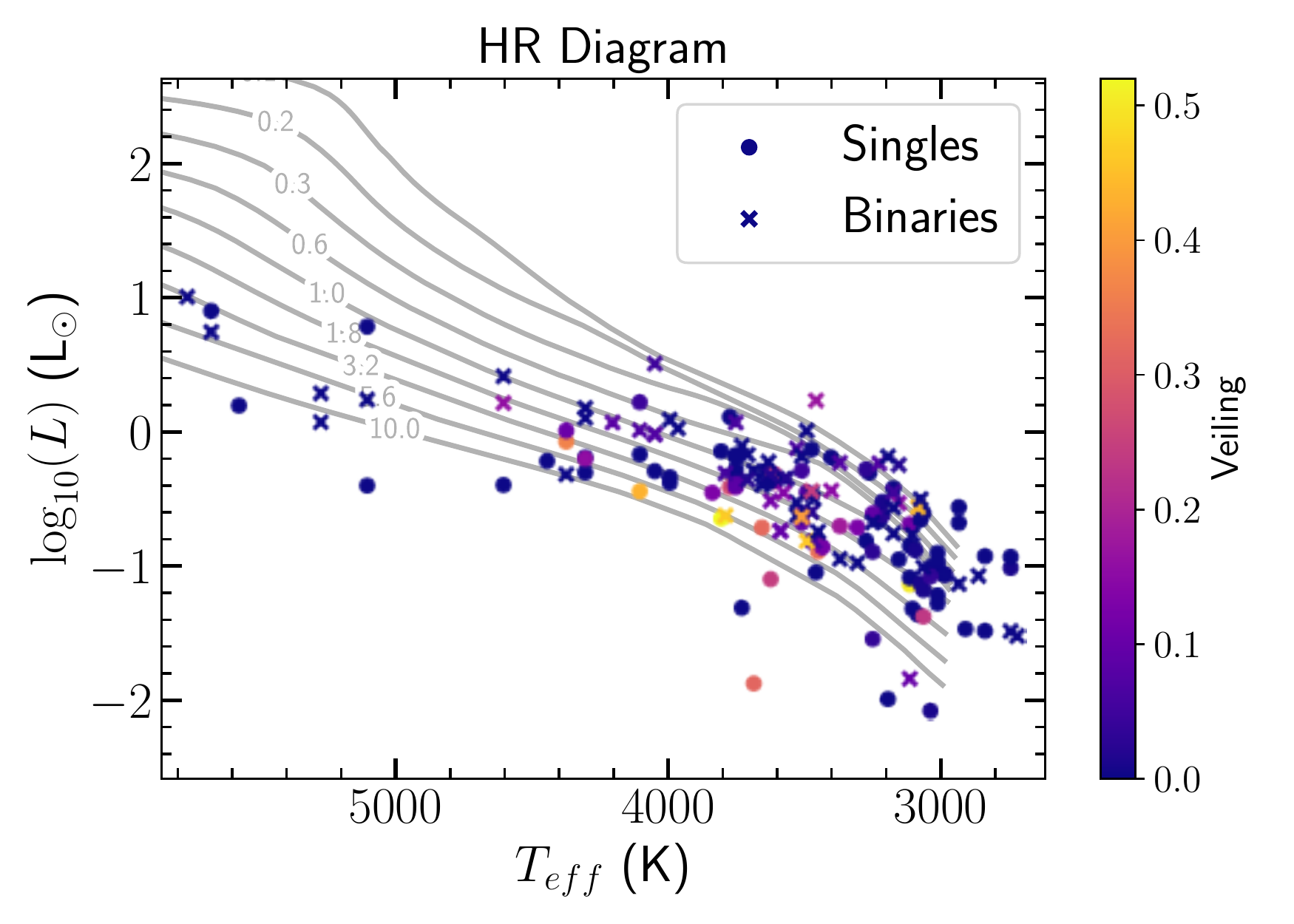}{0.45\linewidth}{}
}
\caption{Top: The observed (left) and de-reddened (right) color-magnitude diagrams plotted as $M_{J}$ vs. 2MASS $J-K$ for the input sample, where the color coding corresponds to the optical veiling value. The arrow on the upper right of each figure denotes a reddening vector for \av = 1 using the \citet{Cardelli1989} reddening law. Binaries are denoted with `x' symbols, while single stars are points. Isochrones from the MIST models spanning ages of 0.1-10 Myr are underlaid on the figures. The majority of points lay to the right of the isochrones, indicating that they are redder than expected due to the presence of a circumstellar disk. The high-veiling stars typically have large $J-K$ excesses. The stars still fall on the red side of the isochrones after being de-reddened because of the circumstellar disks. Bottom: A Hertzsprung-Russell Diagram of the input sample, plotted using the HH14 measured luminosity and a temperature derived by converting the HH14 spectral types to \teff using the \citet{Pecaut2013} SpT-\teff conversion. The color bar again indicates the veiling for each star; binaries are 'x' markers, and single stars are points. Many of the stars are very young, and the high-veiling stars are dispersed among the sample.}
\label{fig:CMD_compare}
\end{figure*}

To select the NIR photometry for our analysis, we crossmatched the HH14 Taurus sample with the 2MASS \citep{Skrutskie2006} and ALLWISE \citep{Wright2010} surveys to find the measured NIR magnitudes ($JHK_{S}W1W2W3W4$) for each target. Although our focus was on the NIR 2MASS filters ($JHK_{s}$), we included the two short-wavelength WISE filters ($\lambda_{eff, W1} = 3.6 \mu$m and $\lambda_{eff, W2} = 4.5 \mu$m) because they are analogous to the L and M bands but did not extend our analysis further into the MIR. However, we used the WISE W4 ($\lambda_{eff, W4} = 22 \mu$m) band to identify disk-bearing stars as described in the following section. After cross-matching with 2MASS and WISE, we retained 163 stars with individual 2MASS and WISE photometry entries corresponding to each optical spectrum from HH14. The $JHK_{S}W1W2$ photometry for each source is listed in Table \ref{tab:source_params}. 

Figure \ref{fig:CMD_compare} shows some of the properties of our sample. The top row shows two color-magnitude diagrams in $M_{J}$ vs. J-K space, with MIST isochrones \citep{Paxton2011, Paxton2013, Paxton2015, Dotter2016, Choi2016} ranging from 0.1 to 10 Myr underlaid in gray. The single stars are marked as points and the binary stars have `x' markers, and all points are color coded by their veiling value. The top left panel shows the systems before correction for extinction, and the top right panel shows the systems after they have been de-reddened. In both panels many systems fall redward of the isochrones because of the presence of a circumstellar disk producing NIR excess. The systems with high veiling typically have large J-K excesses even after being corrected for reddening, indicating that they are disk hosts. 

The bottom panel of Figure \ref{fig:CMD_compare} shows the sample on an HR diagram using the system luminosities from HH14 \edit1{(after rescaling to the \textit{Gaia} distance)}, and a temperature derived from the \citet{Pecaut2013} spectral type-$T_{eff}$ conversion, which we used throughout this work. Ioschrones ranging from 0.1-10 Myr are again underlaid on the figure, and the markers and color scheme are the same as in the top row of the figure. The high-veiling sources are mixed with the remainder of the population. We did not remove the weak-lined T Tauri stars HH14 used as spectral type templates from the sample, so some targets appear much older than others.

\subsection{Identifying Binary Companions}
Multiplicity, especially close stellar companions (separation $\rho < 50$ au, or $\rho < 0.35 \arcsec$ at the typical 145 pc Taurus distance), can greatly complicate disk structure and accretion dynamics \citep[e.g.,][]{Tofflemire2017}. Thus, we wished to identify multiple stars in our sample to explore whether multiplicity impacted the relationship between veiling and NIR excess, and to remove them from our analysis if necessary. 

To identify binaries in our sample, we crossmatched our sample with the binaries identified in high-resolution surveys by \citet{Kraus2011}, \citet{Kraus2012b}, \citet{Schaefer2014}, and \citet{Daemgen2015}. Because we did not search for spectroscopic binaries, our measured multiplicity is a lower limit, but should include the majority of binaries in the sample. We found that 71 of our 163 targets were resolved multiples, and thus measured a multiplicity fraction of $f_{binary} \sim 0.44$, which is slightly lower than the multiplicity fraction of $f_{binary} \sim$ 60\% measured for Taurus \citep{Kraus2011, Duchene2013, Daemgen2015}. This is likely because we had previously removed targets that are known wide-separation binaries with separate spectroscopic observations but unresolved photometry from 2MASS and WISE, which artificially reduced the measured binary fraction. This also meant that our binary sample was comprised mostly of very close ($\rho < 0.5\arcsec$) or very wide ($\rho > 5\arcsec$) binaries. Half of the binary sample (37 out of 71 systems) had a wider separation ($\rho > 0.35\arcsec$, roughly 50 au) than is expected to affect the disk properties. The multiplicity status and source for multiplicity determination of each target is listed in Table \ref{tab:source_params}.

\section{NIR Excess Calculation}\label{sec:excess}
After identifying the 2MASS and WISE magnitudes for each object, we de-reddened the magnitudes by converting the A$_{V}$ measured by HH14 to A$_{\lambda}$ using the reddening law of \citet{Cardelli1989}, which defines $A_{J}/A{V} = 0.282, A_{H}/A_{V} = 0.19, A_{K_{s}}/A_{V} = 0.114$, and $A_{W1}/A_{V} = 0.056$. In W2, W3, and W4, where the \citet{Cardelli1989} extinction law was not defined, we assumed the extinction coefficient was effectively zero. Although there are other extinction law determinations that extend further into the infrared \citep[e.g.,][]{Davenport2014}, we used the \citet{Cardelli1989} law to match HH14.

To calculate the intrinsic magnitudes for our sample, we began with the bolometric luminosity values measured by HH14 from their flux-calibrated spectra. We converted $L_{bol}$ to an absolute bolometric magnitude by asserting that $M_{bol} = -2.5\log_{10}(L_{bol}/L_{\odot}) + M_{bol, \odot}$, where $M_{bol, \odot}$ = 4.74. \edit1{Using the appropriate distance for each system, we converted the absolute bolometric magnitude $M_{bol}$ to an apparent magnitude $m_{bol}$.}


\begin{figure}
\gridline{\fig{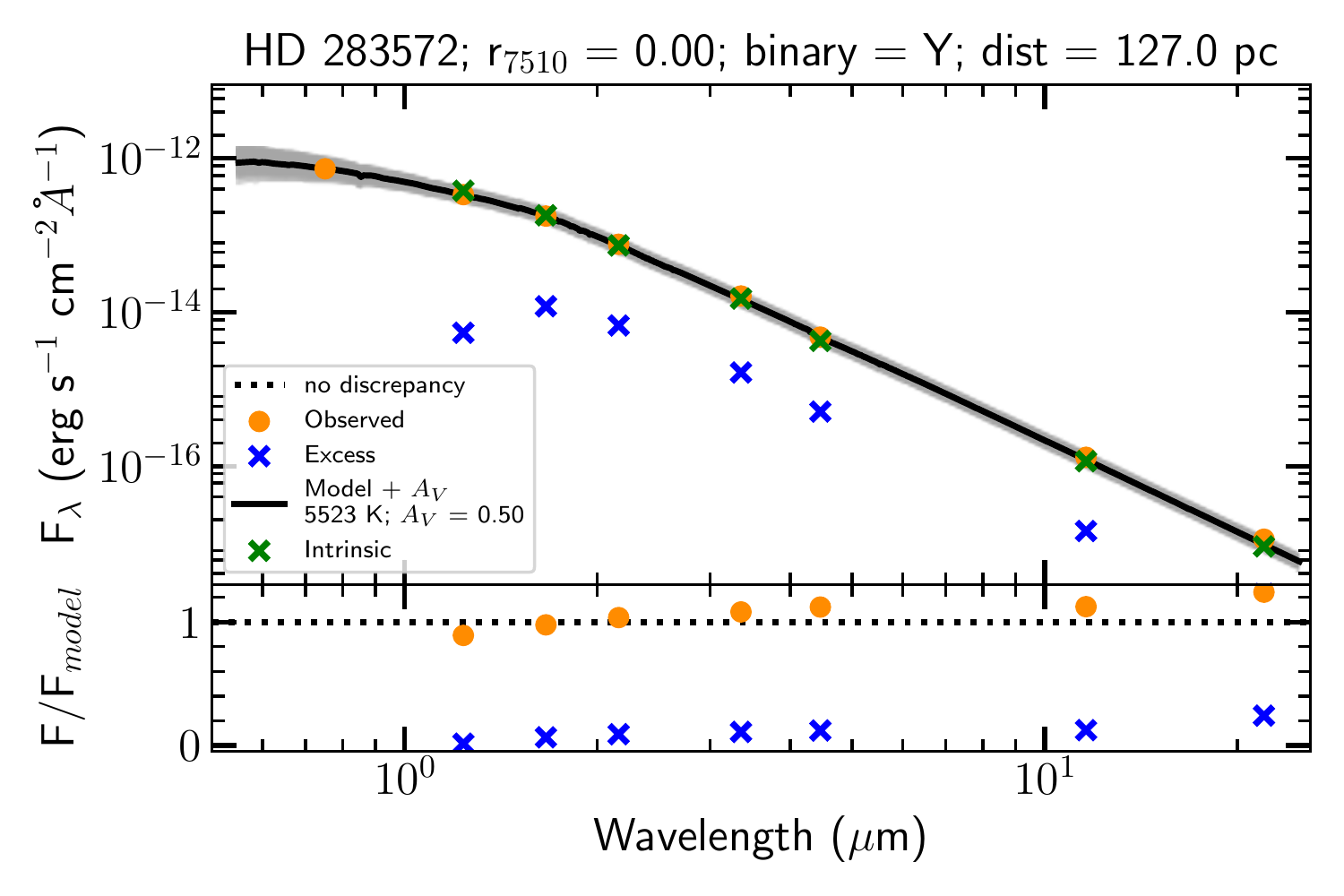}{0.95\linewidth}{}
	}
\gridline{\fig{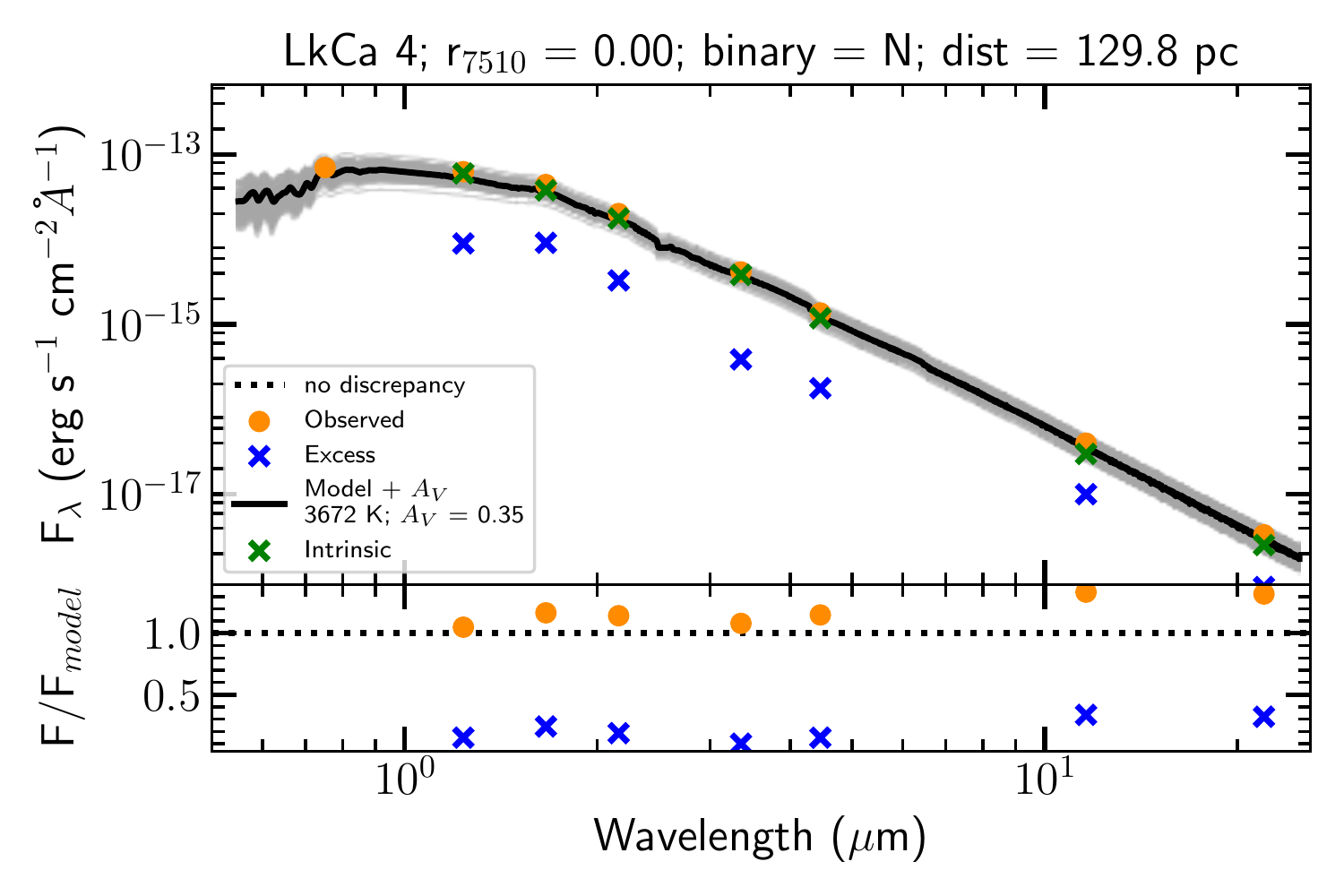}{0.95\linewidth}{}
}
\gridline{\fig{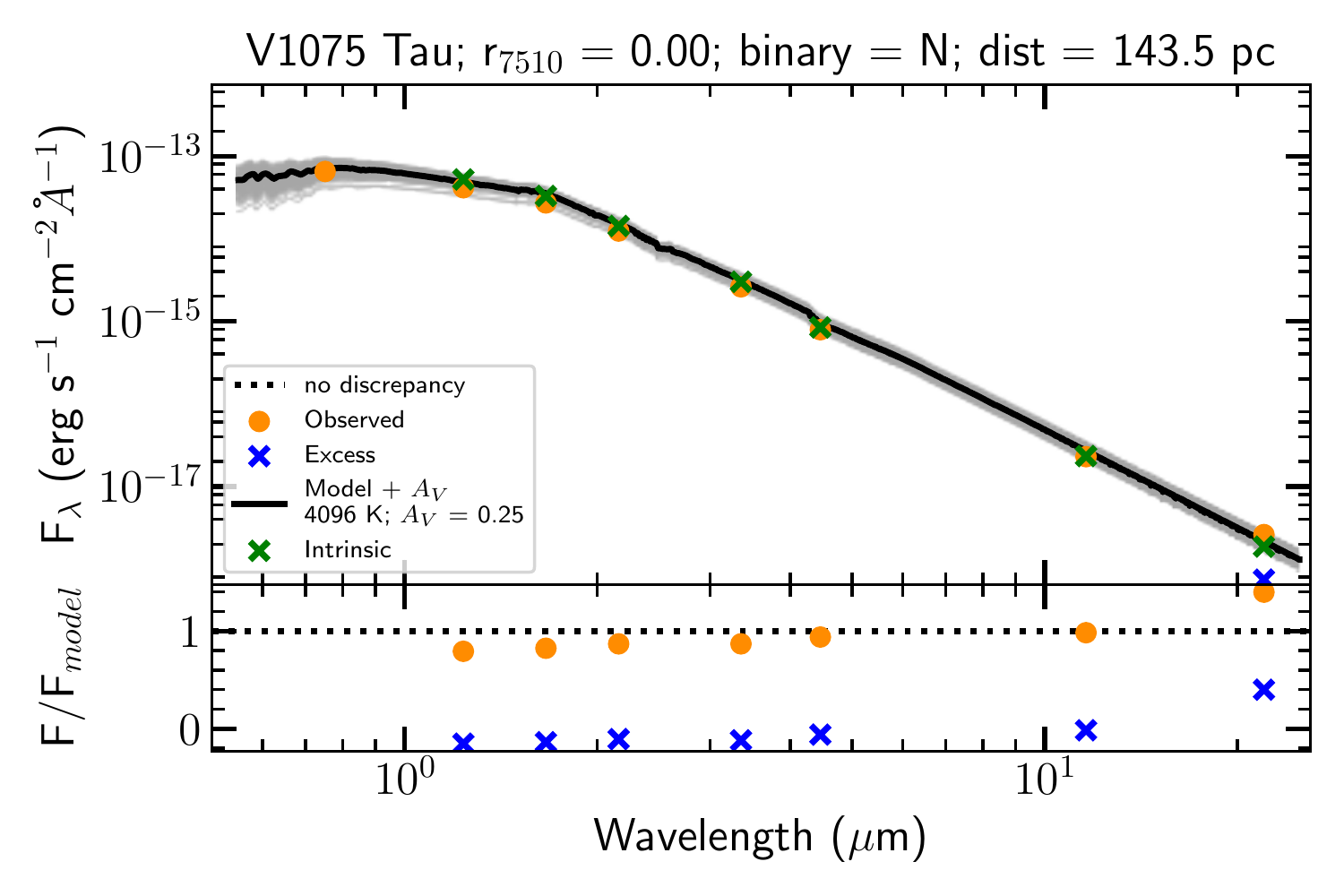}{0.95\linewidth}{}
}
\caption{Infrared SEDs for several example systems without disks. The markers and color scheme are the same as in Figure \ref{fig:source_SEDs}. These stars do not have a significant IR excess. The remainder of the disk-free SEDs are shown in a Figure Set in the online version of this paper.} 
\label{fig:nodisk_source_SEDs}
\end{figure}

We converted each $m_{bol}$ to a J magnitude using the \citet{Pecaut2013} J-band bolometric correction as $J = m_{bol} - BC_{J}$. Then, we inferred the remaining predicted photospheric magnitudes ($HK_{s}W1W2W3W4$) using the appropriate intrinsic colors from \citet{Pecaut2013}. Finally, we subtracted the predicted intrinsic magnitudes from the de-reddened observed magnitudes to measure the excess in each photometric band. 

Figure \ref{fig:nodisk_source_SEDs} shows spectral energy distributions (SEDs) of several systems without disks, including the observed system fluxes, the intrinsic stellar fluxes, and the excess (disk) fluxes, which for these disk-free systems are typically close to zero. To demonstrate the error on the SED calculation caused by observational uncertainties in \av, bolometric luminosity, and spectral type, we also plotted 100 samples of a BT-Settl model spectrum \citep{Allard2003, Barber2006, Allard2011, Caffau2011, Allard2012, Allard2013} calculated with perturbed values of those parameters drawn from a normal distribution with a standard deviation equal to the quoted measurement error from HH14. From visual inspection of all system SEDs we identified nine sources with incorrect \av\ measurements and two systems (DG Tau and IQ Tau) with otherwise peculiar SEDs. We removed these 11 systems from our analysis, because the SEDs indicated that we could not calculate an accurate NIR excess. Systems that were removed from the analysis because of anomalous SEDs are marked with a note in Table \ref{tab:source_params}. Figure Set 1 in the online version of this paper shows SEDs for all the disk-free systems in our sample.

The measurement error on \av, L$_{bol}$, and SpT also affects the inferred NIR excess by altering the predicted intrinsic magnitude for a system. To assess the error on the NIR excess we used a bootstrap analysis. Using 1000 samples for each system, we randomly drew a spectral type, luminosity, and extinction from a normal distribution with a mean equal to the published value for each system and a FWHM equal to the error on the measurement given by HH14: $\sigma_{SpT} = 0.5$ subclass; $\sigma_{L_{bol}} = 10\%$; $\sigma_{A_{V}} = 0.3$mag. Using the perturbed values, we inferred a NIR excess using the procedure described above, and took the error on each excess to be the standard deviation of the 1000 bootstrapped draws. To calculate the error on the color excess we added the bootstrapped errors for the two relevant NIR bands in quadrature.

\subsection{Infrared Disk Indicators}
As part of our sample selection process we wished to only select sources with disks, as those are the systems that should be accreting significantly. To identify disks we could not simply select sources with nonzero veiling, because young stars have variable accretion, meaning that stars with disks may be in a quiescent accretion phase and thus have no measured veiling. A more reliable indicator of disk presence is mid-IR excess, usually characterized by a K$_{s} - [24]$ or [24] $\mu$m excess, where [24] is the Spitzer 24 $\mu$m band.


\begin{figure*}
\gridline{\fig{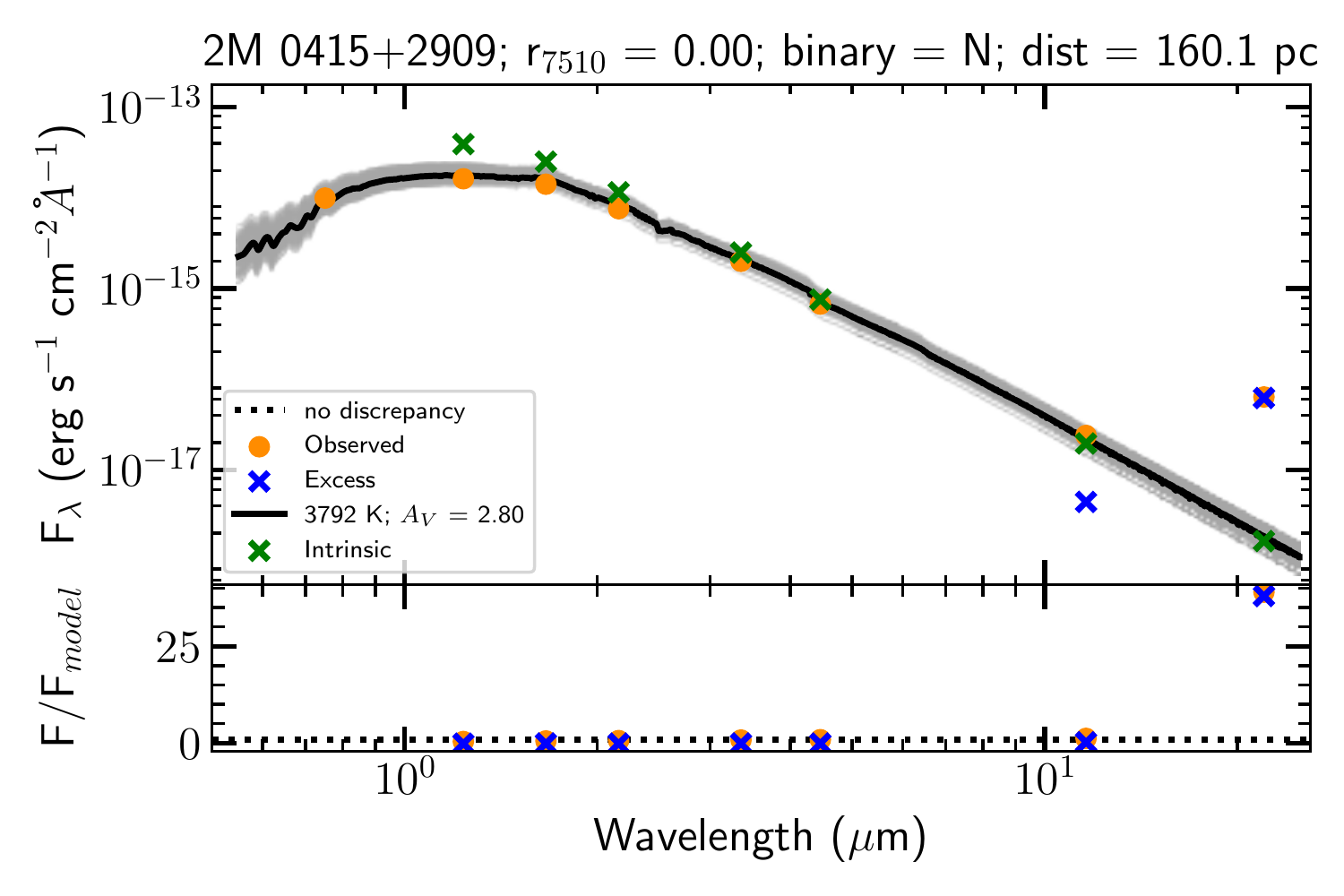}{0.45\linewidth}{}
    \fig{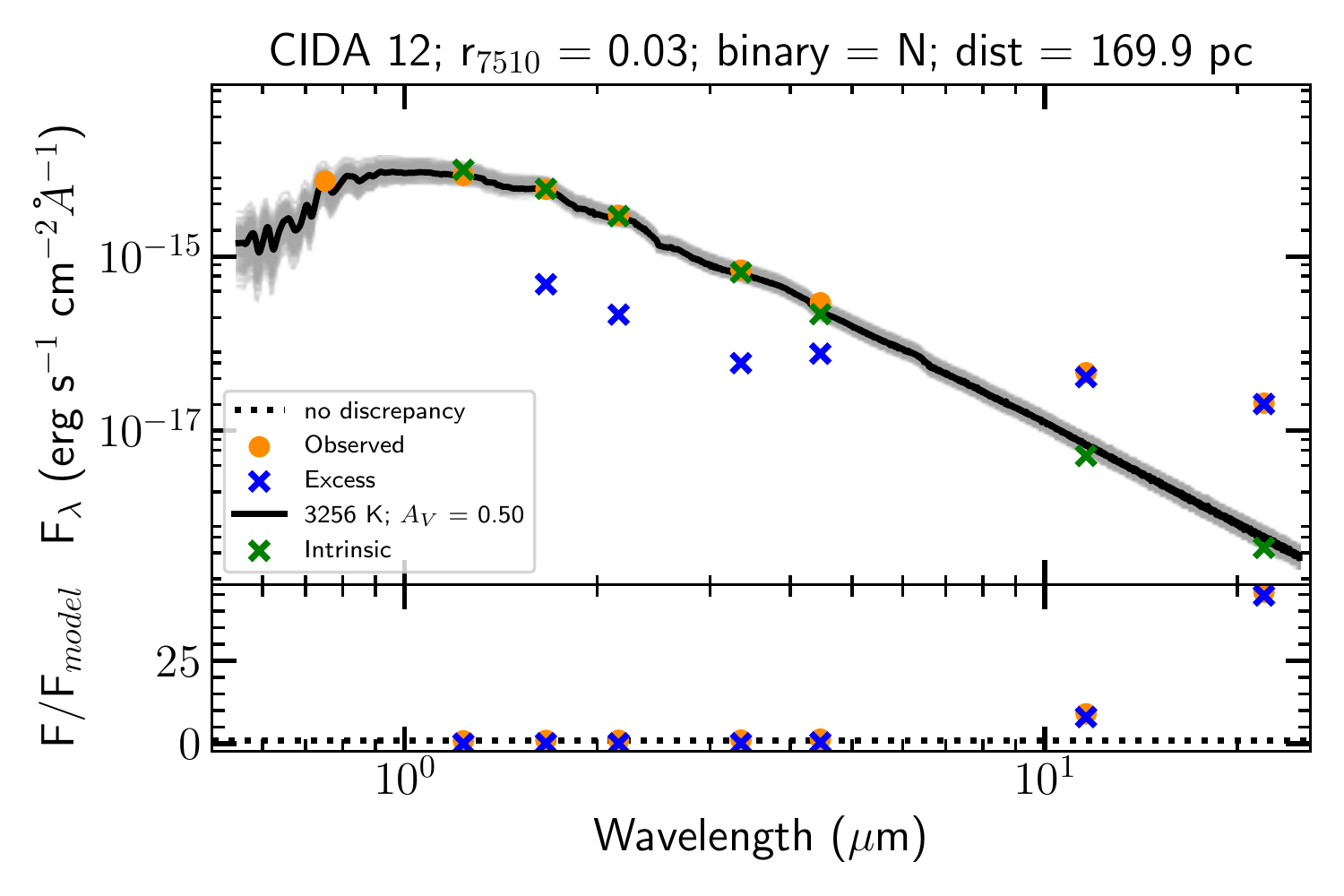}{0.45\linewidth}{}
	}
\gridline{\fig{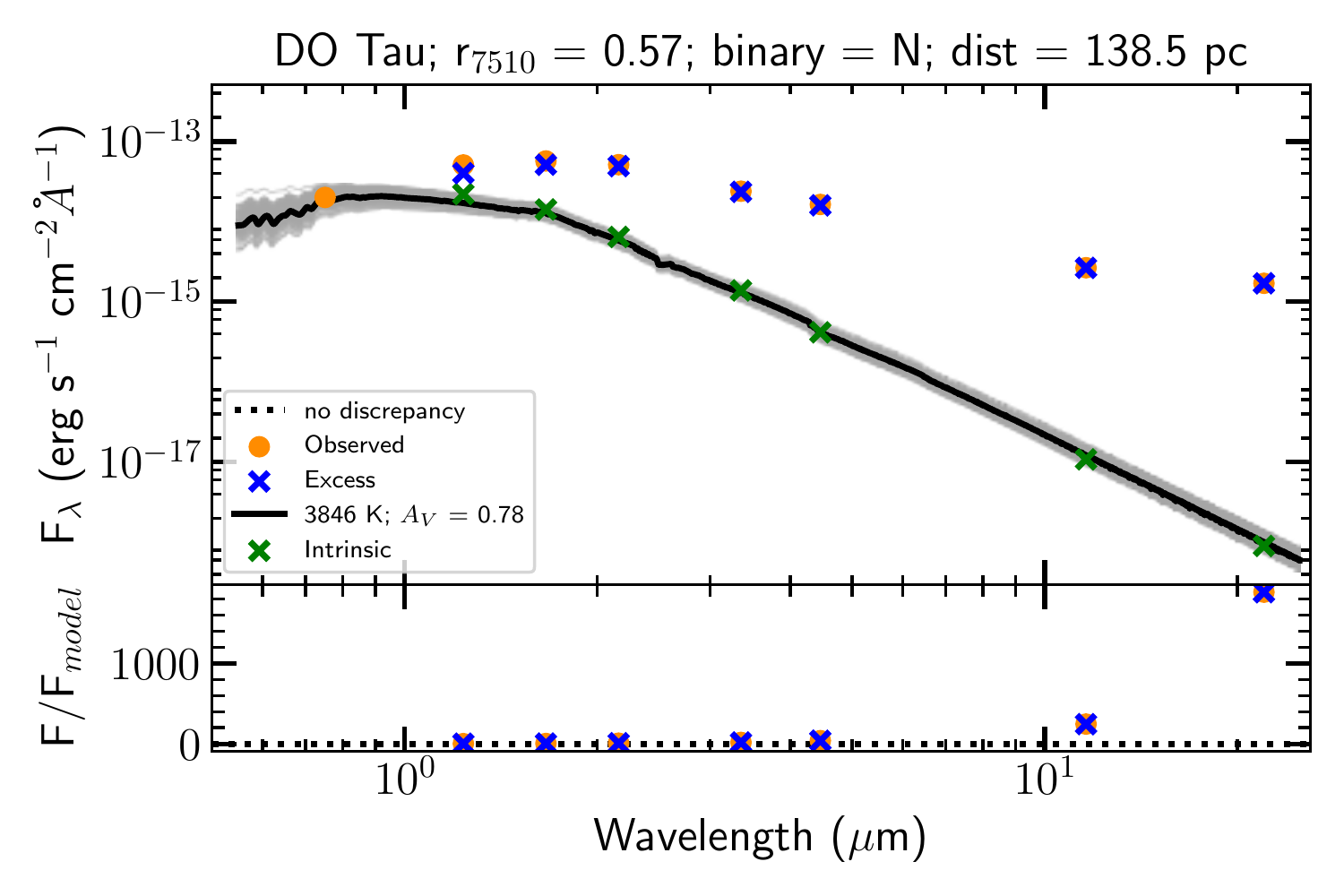}{0.45\linewidth}{}
        \fig{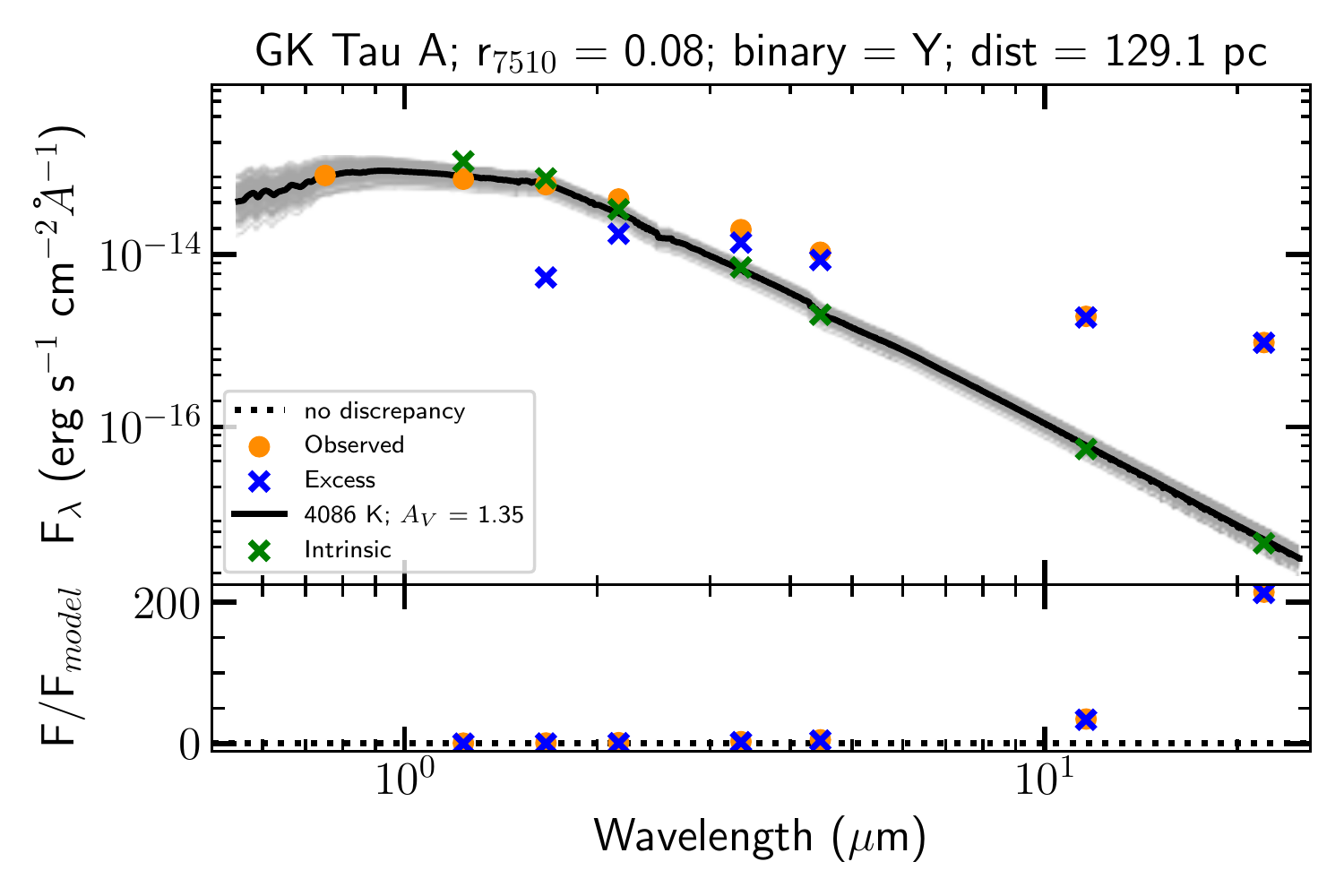}{0.45\linewidth}{}
}
\gridline{\fig{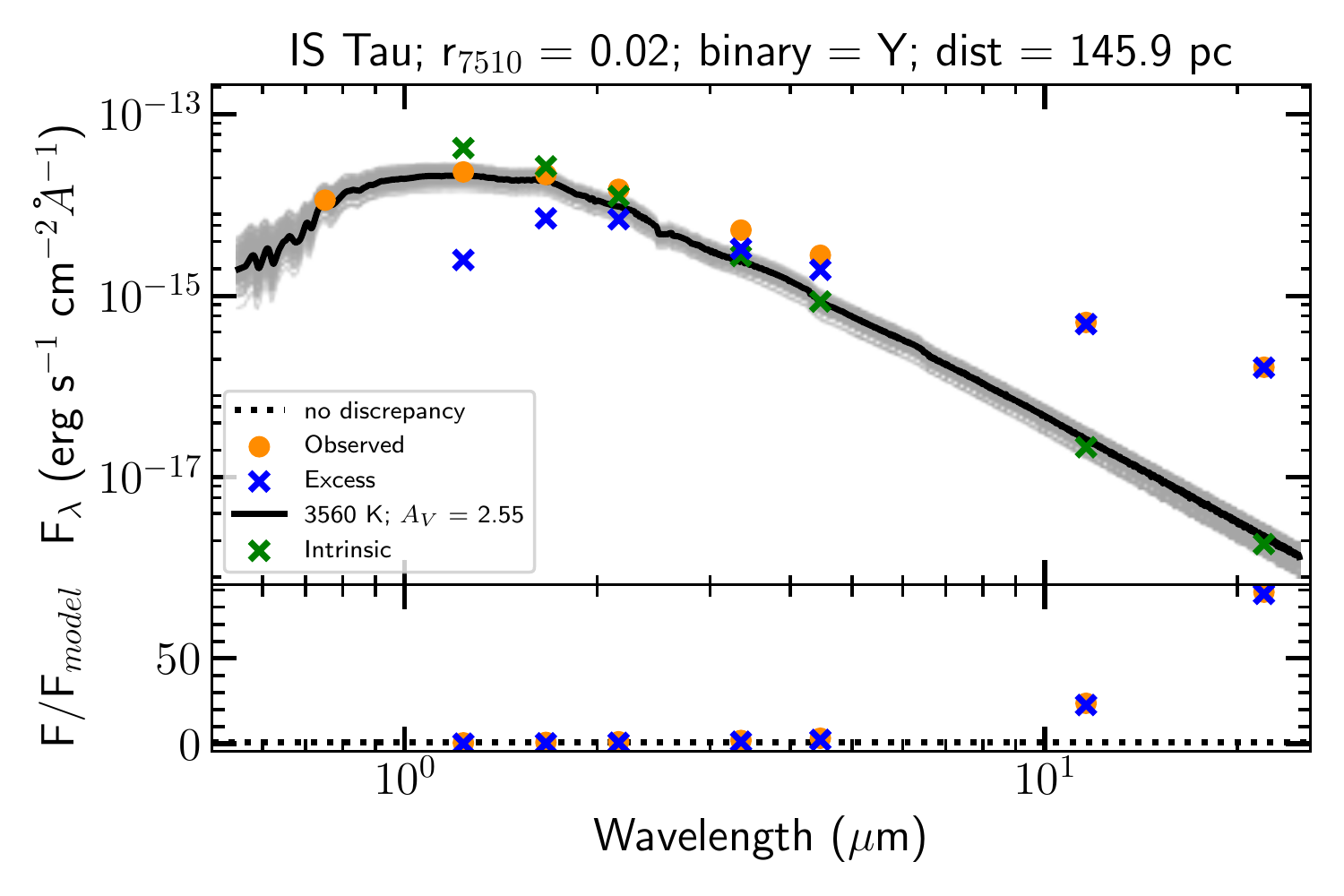}{0.45\linewidth}{}
        \fig{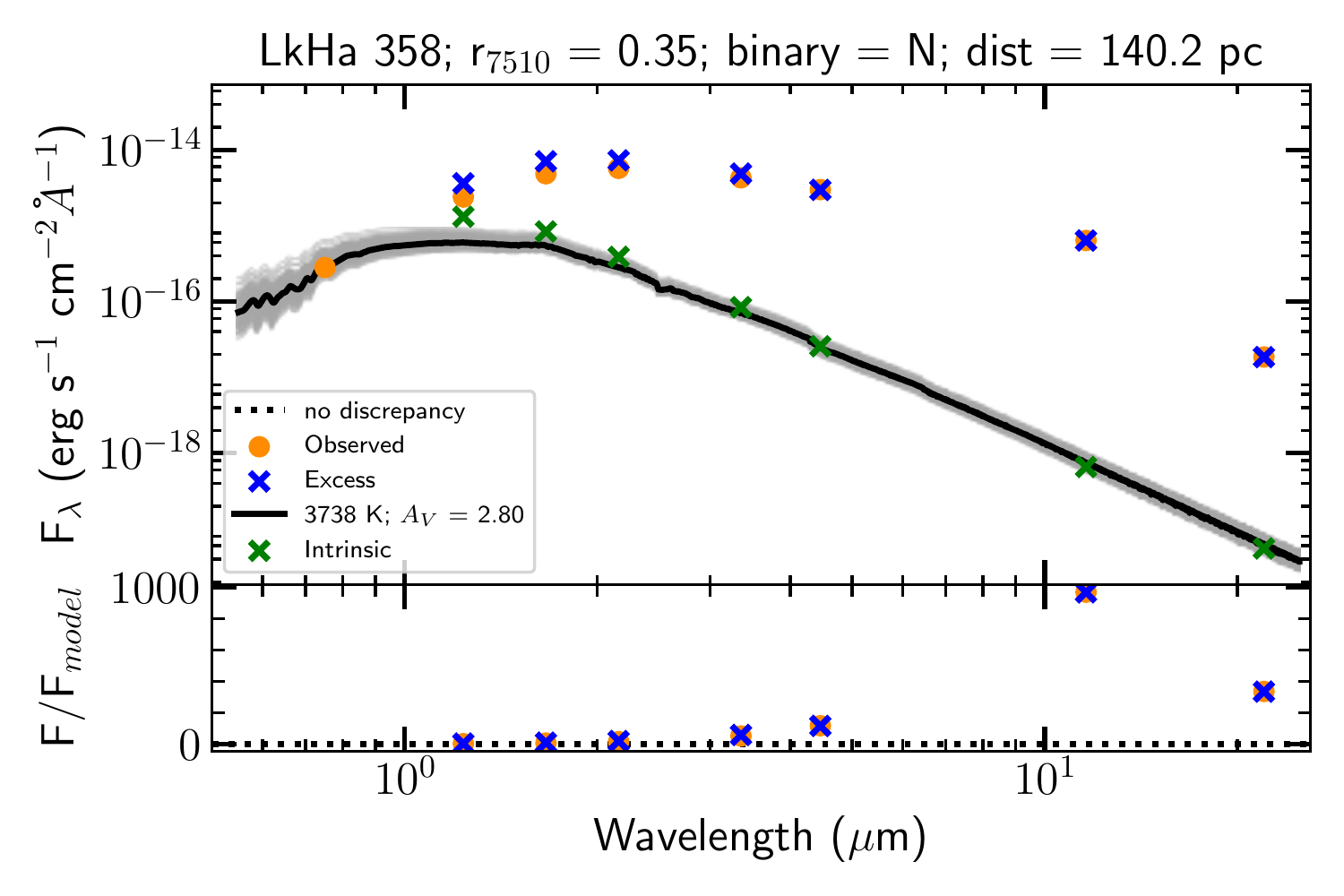}{0.45\linewidth}{}
}
\caption{Infrared SEDs for several example systems with disks. A BT-Settl model spectrum of the appropriate temperature, with solar metallicity and surface gravity of $\log(g) = 4.5$, is plotted in black, with the results from 100 MCMC draws shown in gray. The observed (orange points), intrinsic (green `x'), and excess (blue `x') fluxes are plotted in units of erg s$^{-1}$ cm$^{-2}$ \AA$^{-1}$. The example spectrum has fluxes calculated using the \edit1{HH14 temperature, \textit{Gaia} distance, and rescaled L$_{bol}$}. The orange F$_{7510}$ point was measured by HH14 and corrected for excess/veiling, and extincted in this work using the measured A$_{V}$ from HH14 and the \citet{Cardelli1989} extinction law, with R$_{V}$ = 3.1}. The remainder of the disk-host star SEDs are shown in a Figure Set in the online version of this paper.
\label{fig:source_SEDs}
\end{figure*}

We used two methods to identify disk-bearing stars in our sample. First, we assumed that the WISE W4 band, with $\lambda_{cen} = 22 \mu$m, was analogous to the Spitzer 24$\mu$m band. We used a disk excess metric from \citet{Rebull2010}, which found that stars with disks have $K_{s} < 14$ and $K_{s} - [24] > 1$, where we replaced [24] with W4. Using this metric, we identified 133 stars with disks. We crossmatched the remaining 28 stars with the \citet{Rebull2010} Spitzer survey of Taurus, since Spitzer was more sensitive at 24 $\mu$m than WISE was at 22 $\mu$m, but did not find any additional disk-hosting stars. Thus, the final sample of disk-bearing stars consisted of 133 systems. The disk host status of each target is listed in Table \ref{tab:source_params}. 

Figure \ref{fig:nodisk_source_SEDs} shows example SEDs for three systems that do not have an IR disk excess, while Figure \ref{fig:source_SEDs} shows SEDs of several representative stars that host disks. SEDs for all stars in the sample are contained in Figure Sets 1 and 2 in the online version of this paper. The disk-free stars are fit well by both the intrinsic colors calculated using \citet{Pecaut2013} and the model spectra calculated using the BT-Settl models, and show little to no excess across the full wavelength range of the SED. In contrast, the disk host star SEDs all show excess that is orders of magnitude larger than the intrinsic stellar photospheric flux at long wavelengths, and often have significant excesses, with fluxes comparable to the intrinsic stellar brightness, across the full SED. 

\section{Correlations Between Optical Veiling and NIR Excesses} \label{sec:veiling correlations}

\subsection{Relationships Between Veiling and NIR Colors}
Any correlation between optical veiling and NIR excess is driven by the relation between accretion in two different regimes of the star-disk system. The accretion shock on the stellar surface produces UV and optical veiling, so changes in optical veiling should be caused by changes in accretion. In the NIR, the emission is (at least predominantly) caused by the inner rim of the dust disk, and changes to the environment of the inner disk, such as by accretion variability, should cause changes in the disk emission. Thus, quantifying and understanding the relationship between these two seemingly disparate processes informs our understanding of accretion and its potential effects on the circumstellar disk environment. The most easily observable relations are the correlations between IR color excesses and veiling, which were the focus of past work \citep[e.g.,][]{Hartigan1990, Kenyon1995}. Thus, we began by exploring those correlations in the NIR for our sample. 

The color excess describes how red a source is by measuring the flux ratio between two points of its SED. A larger color excess indicates a larger flux ratio, which could be produced by either a larger absolute excess or a steeper SED slope, either of which is independent of the disk luminosity because color is a relative measure. In addition, color excess is not directly dependent on the stellar luminosity, because it is a relative measurement, rather than one that relies on removing the underlying stellar flux. To quantify the relationship between NIR color excess and optical veiling, we calculated linear regressions for the color excess as a function of the optical veiling $r_{7510}$. To explore any possible differences between single and binary star relations, we calculated separate regressions for each subsample of stars with disks. 

We also calculated separate regressions for samples including or excluding the data points where $r_{7510} = 0$. The sample that includes the points with $r_{7510} = 0$ encompasses the range of possible outcomes for disk-bearing stars, from (nearly) quiescent accretion to accretion luminosity nearly comparable to the stellar photosphere. However, we found that the linear regression for that sample had a slope that was driven by stars with $r_{7510} = 0$, and hence with quiescent accretion. Therefore, we also wished to investigate the relationship between veiling and NIR color excess for only actively accreting systems.

\begin{figure*}
\gridline{\fig{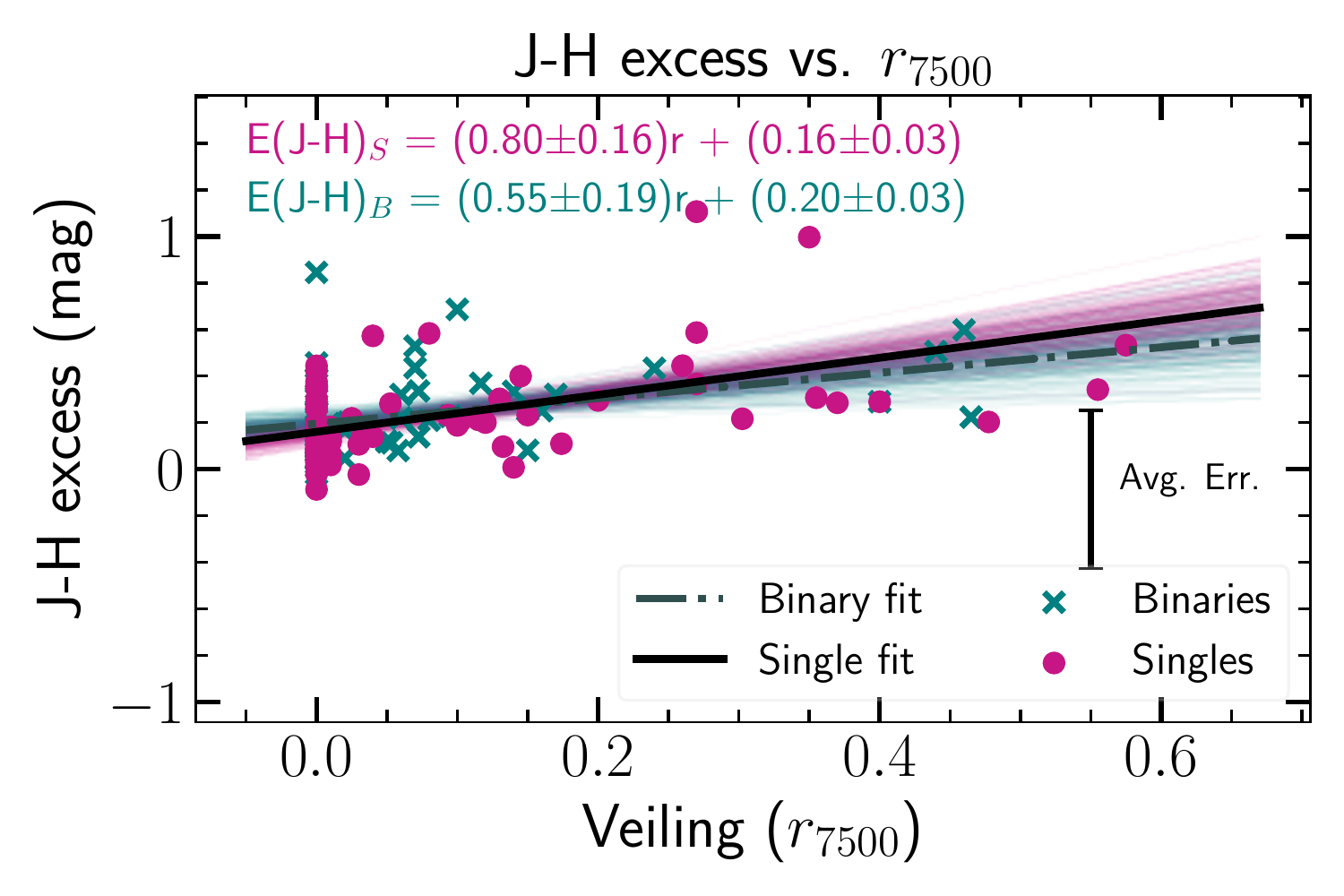}{0.45\linewidth}{}
 	\fig{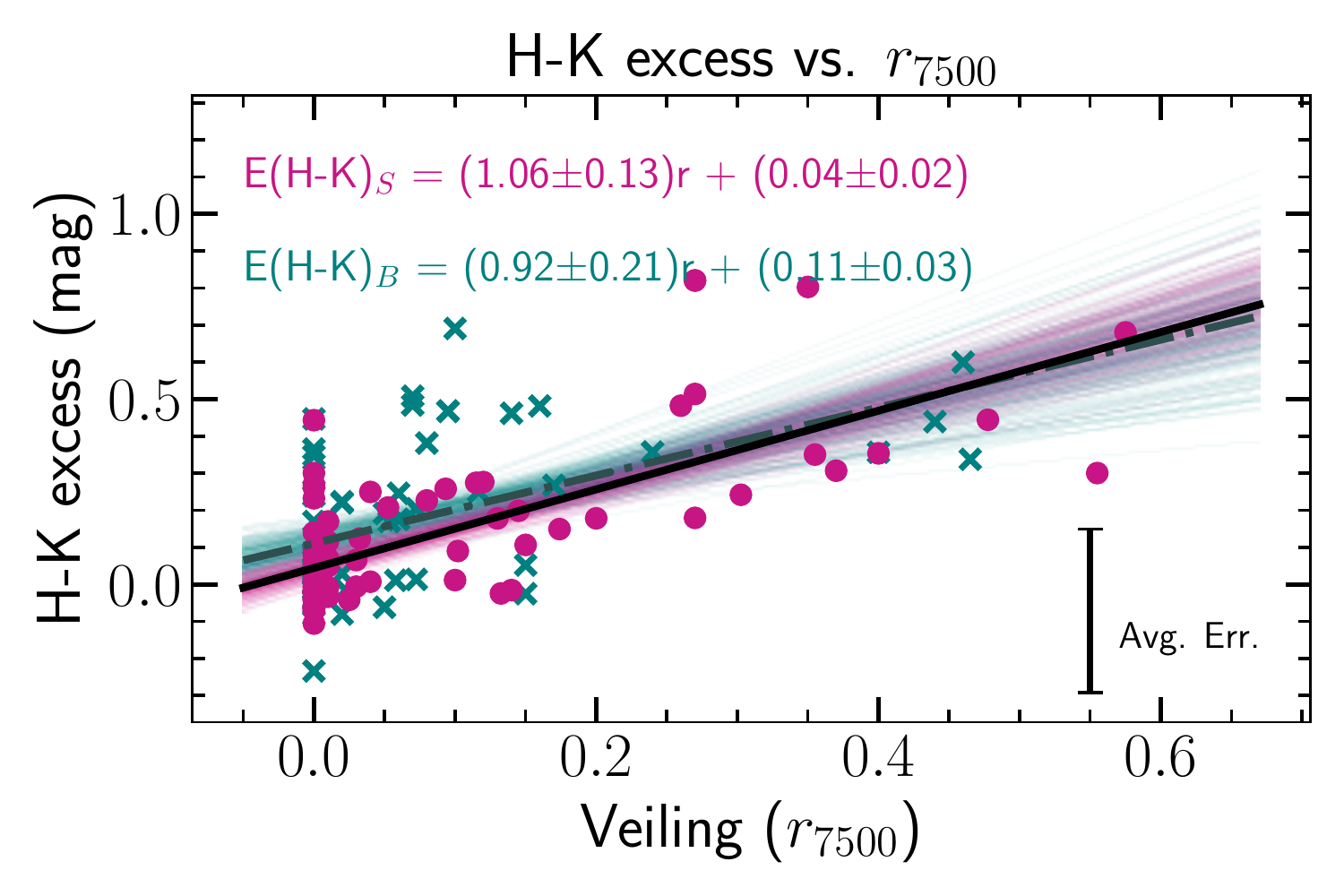}{0.45\linewidth}{}
	}
 \gridline{\fig{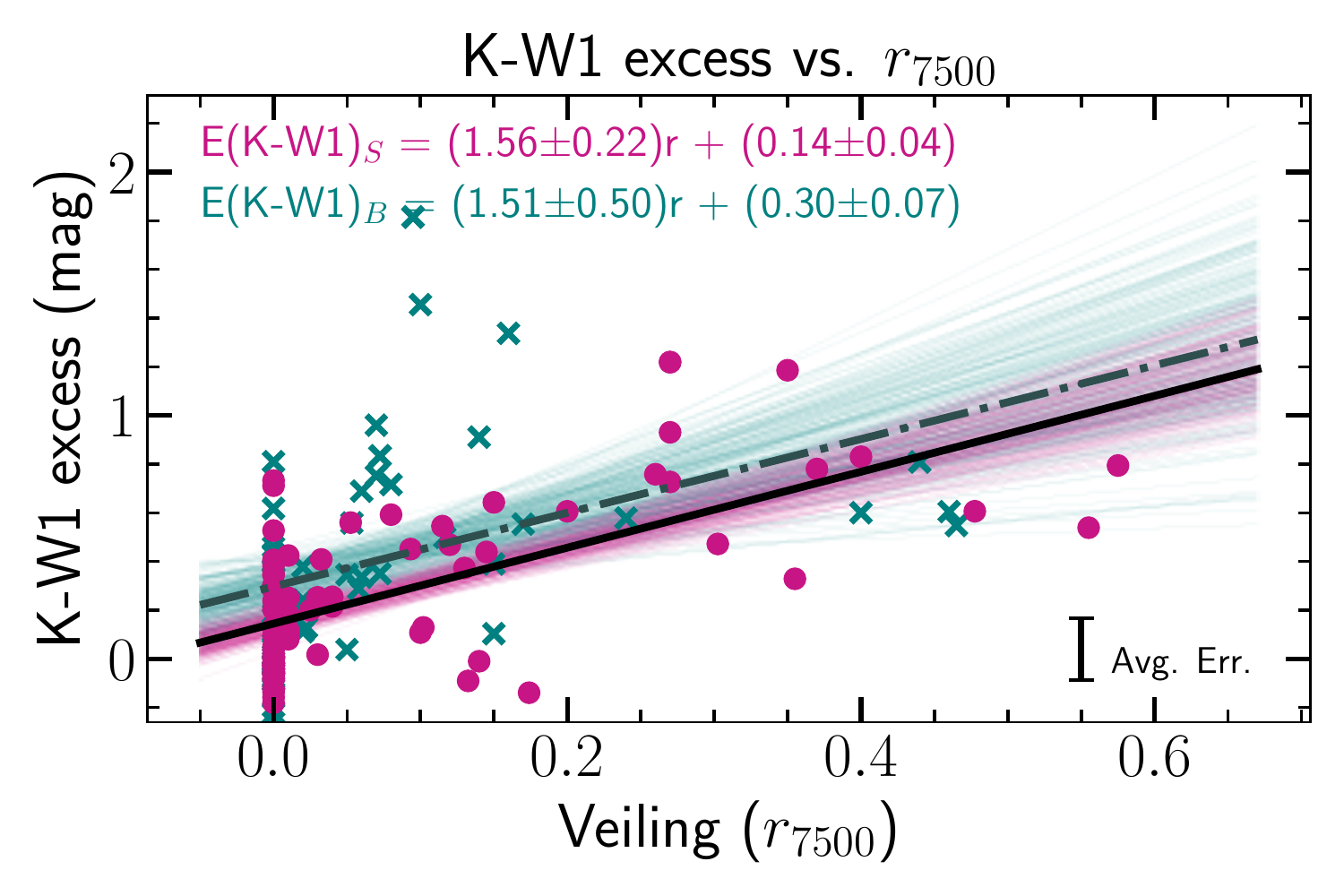}{0.45\linewidth}{}
 	\fig{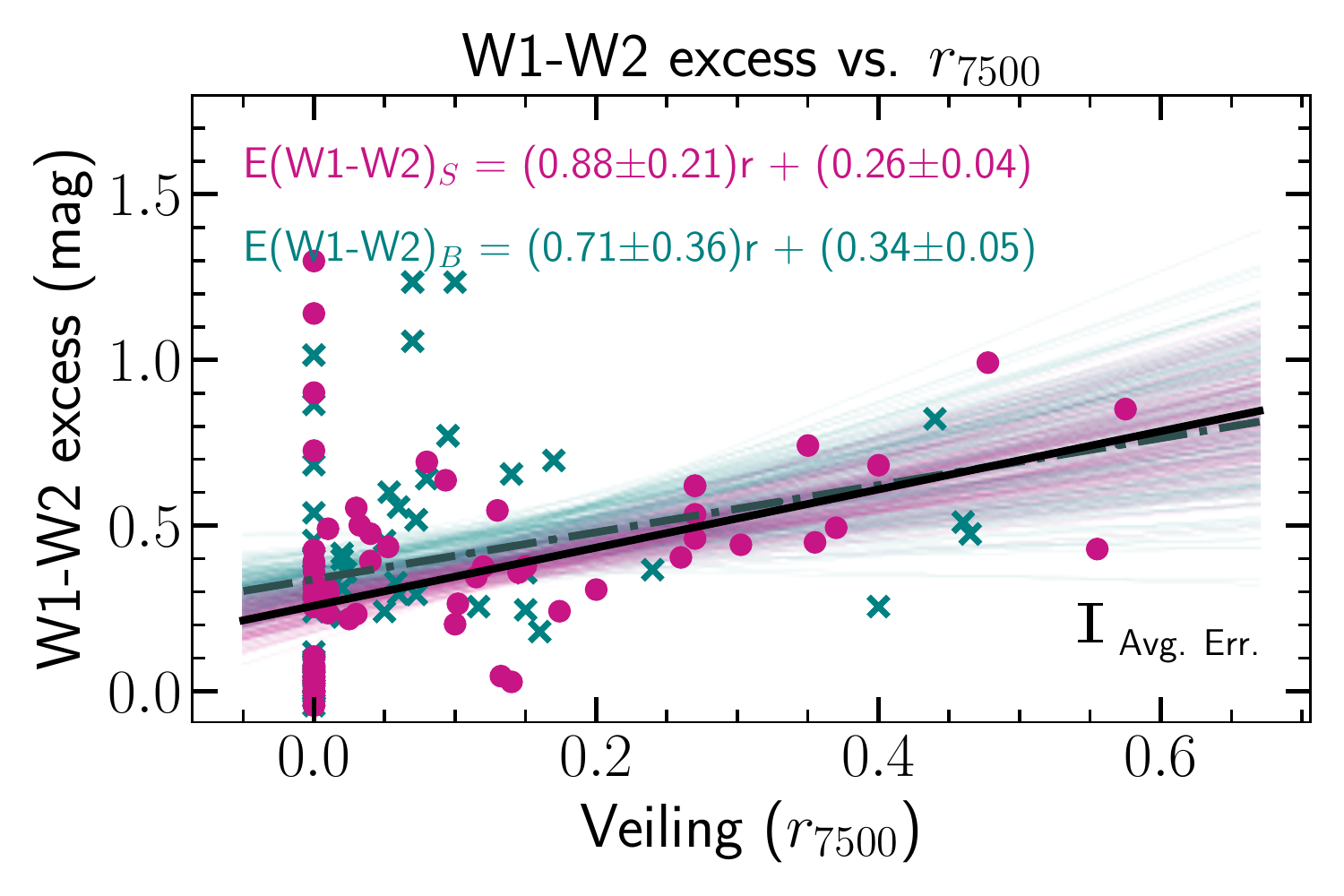}{0.45\linewidth}{}
 	}
\caption{Veiling-excess correlations for NIR colors including points where $r_{7510} = 0$. The binaries are shown in teal with `x' markers, and the single stars are shown as magenta dots. The best-fit lines for the binary and single star regressions are shown as a dot-dash dark blue and solid black line, respectively. The magenta and teal lines underlaid on the figure indicate the error on the best-fit line, calculated using 200 bootstrapped values for the slope and y-intercept of each line. The magenta and teal text on each figure is the equation of the best-fit line for the single and binary stars, respectively, calculated from a linear regression. The binary star slopes are shallower than the single star slopes, indicating that the binary star SEDs are not as red (e.g., not as steeply sloped) as the single stars. }
\label{fig:veiling_color_corr}
\end{figure*}


\begin{figure*}
    \plottwo{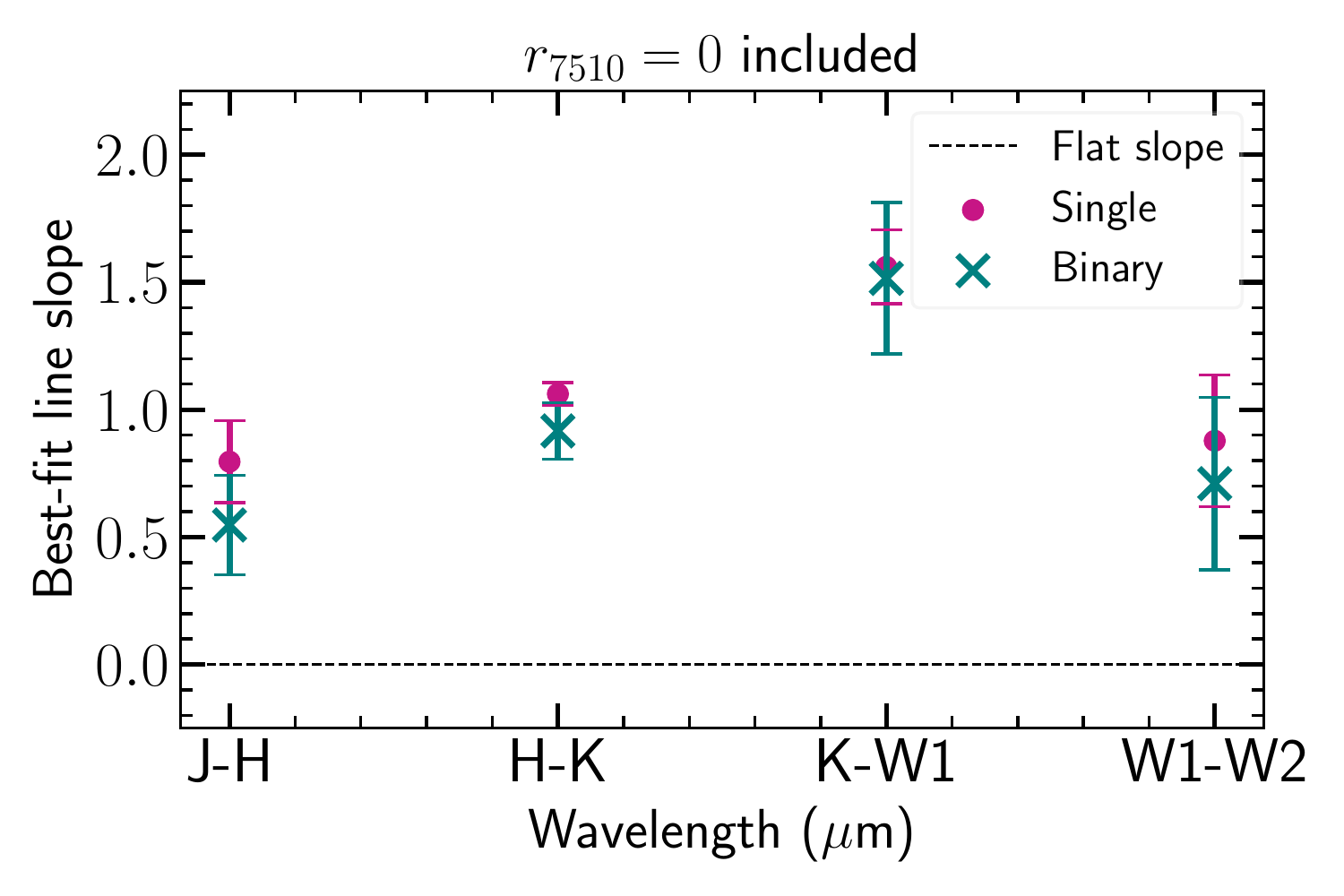}{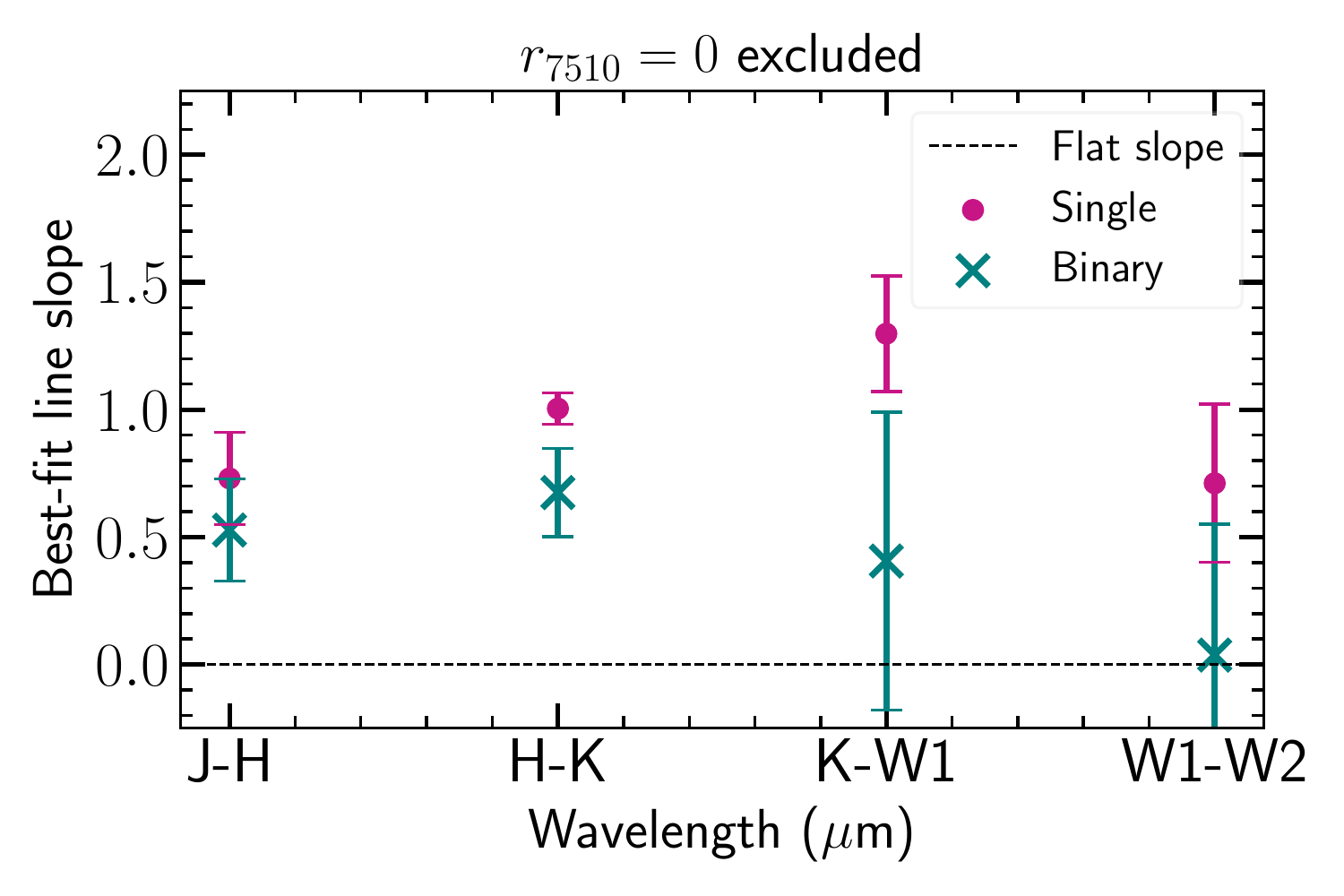}
    \caption{The slopes of the best fit line relating the optical veiling and color excess for binary and single stars (teal and magenta, respectively), where the error bars are the RMS error on the slope, with (left) and without (right) systems with r$_{7510} = 0$. \textbf{Left:} There are no statistically significant differences between the single and binary slopes for the regression including quiescent disk hosts (systems where $r_{7510} = 0$). \textbf{Right:} There are significant differences between the single and binary slopes for \edit1{H-K and K-W1}. When excluding the $r_{7510} = 0$ points, the binary slopes are shallower than when the $r_{7510} = 0$ points are included in the regression.}
    \label{fig:color_slopes}
\end{figure*}

Figure \ref{fig:veiling_color_corr} shows plots of color excess versus $r_{7510}$ for each filter combination including points where $r_{7510} = 0$. For each color we fit a linear trend to the binary and single stars separately. The best-fit lines from each linear regression are plotted in Figure \ref{fig:veiling_color_corr}, along with the data points used in the calculation. The thin magenta and teal lines indicate the error on the linear fit, visualized by 200 bootstrapped lines with parameters \edit1{drawn using the covariance matrix from the linear fit}. We found that almost all of the linear regressions (except the binary W1-W2 relation) were statistically significant and consistent with a non-zero slope (p-value $<$ 0.05), indicating that there is a correlation between color excess and optical veiling in J-H, H-K, and K-W1 colors, as well as W1-W2 for single stars. There is no significant difference in slope between binary and single stars for any color in the regressions that include $r_{7510} = 0$ points, as illustrated by the left panel of Figure \ref{fig:color_slopes}.

After excluding $r_{7510} = 0$ points, all the measured slopes are shallower. The discrepancy between the single and binary stars grows moving redward, and H-K and K-W1 have slopes that differ more than the mutual uncertainty. Shallower slopes in $r_{7510}$ vs. color excess space indicate that the slope of the mean binary SED is shallower than the mean of the slope of the single star SED, especially at longer wavelengths in actively-accreting systems.

\begin{deluxetable*}{cCCCCCCC}
\tablecaption{Best-fit line parameters, P-values, and $\rho$ values for each NIR color \label{tab:color_params}}
\setcounter{table}{1}
\tablecolumns{8}
\tablewidth{0pt}
\tablehead{
\colhead{Color} & \colhead{Slope} & \colhead{Y-intercept}  & \colhead{P-value} & \colhead{Spearman $\rho$} & \colhead{Slope} & \colhead{Y-intercept} & \colhead{Y-intercept}\\
\colhead{} & \colhead{$f(r_{7510})$} & \colhead{$f(r_{7510})$} & \colhead{$f(r_{7510})$} & \colhead{$f(r_{7510})$} & \colhead{(no $r=0$)} & \colhead{(no $r=0$)} & \colhead{$f(r_{7510}-\overline{r_{7510}})$\tablenotemark{a}}\\
}
\startdata
\multicolumn{8}{c}{Single Stars}\\
\hline
J-H & 0.80 $\pm$ 0.16 & 0.16$\pm$ 0.03  & $<0.0001$ & 0.49 & 0.73 $\pm$ 0.25 & 0.18$\pm$ 0.06 & 0.23$\pm$ 0.04\\
H-K & 1.06 $\pm$ 0.13 & 0.04$\pm$ 0.02 & $<0.0001$ & 0.66 & 1.00 $\pm$ 0.18 & 0.06$\pm$ 0.04 & 0.12$\pm$ 0.03\\
K-W1 & 1.56 $\pm$ 0.22 & 0.14$\pm$ 0.04 & $<0.0001$ & 0.63 & 1.30 $\pm$ 0.29 & 0.23$\pm$ 0.07 & 0.24$\pm$ 0.05\\
W1-W2 & 0.88 $\pm$ 0.21 & 0.26$\pm$ 0.04 & $<0.0001$ & 0.54 & 0.71 $\pm$ 0.17 & 0.31$\pm$ 0.04 & 0.33$\pm$ 0.04\\
\hline
\multicolumn{8}{c}{Binary Stars}\\
\hline
J-H & 0.55 $\pm$ 0.19 & 0.20$\pm$ 0.03 & 0.001 & 0.42 & 0.53 $\pm$ 0.20 & 0.20$\pm$ 0.04 & 0.24$\pm$ 0.03\\
H-K & 0.92 $\pm$ 0.21 & 0.11$\pm$ 0.03 & $<0.0001$ & 0.56 & 0.67 $\pm$ 0.26 & 0.17$\pm$ 0.05 & 0.22$\pm$ 0.04\\
K-W1 & 1.51 $\pm$ 0.50 & 0.30$\pm$ 0.07 & $<0.0001$ & 0.68 & 0.41 $\pm$ 0.64 & 0.58$\pm$ 0.12 & 0.61$\pm$ 0.09\\
W1-W2 & 0.71 $\pm$ 0.36 & 0.34$\pm$ 0.05 & 0.0001 & 0.49 & 0.04 $\pm$ 0.40 & 0.51$\pm$ 0.07 & 0.52$\pm$ 0.06\\
\enddata
\tablenotetext{a}{The y-intercept of the line calculated using a regression with the x-values centered at the mean of the veiling values, which is $\overline{r_{7510}} = 0.072$.}
\end{deluxetable*}

The numerical values (slope, Y-intercept, and P-value) of each linear regression are listed in Table \ref{tab:color_params}, and the equation of the best-fit line is also shown on each panel of Figure \ref{fig:veiling_color_corr} for binaries and single stars. Table \ref{tab:color_params} presents \edit1{three} values for the linear fit y-intercept \edit1{and two slope values}. The first value (denoted as $f(r_{7510})$) was calculated using the data without any modification. However, this calculation introduces a covariance between the slope and y-intercept errors, artificially reducing the measured error on the y-intercept. Thus, we also calculated the y-intercept and its associated error after shifting the data so that the veiling values were centered at 0 (i.e., after subtracting the mean veiling value $\overline{r_{7510}} = 0.072$ from all veiling measurements). This additional y-intercept and associated error are denoted as $f(r_{7510} - \overline{r_{7510}})$ in Table \ref{tab:color_params}. \edit1{Finally, we also show the slope and y-intercept calculated when performing the linear fit while excluding the $r_{7510} = 0$ points.}

The p-values found via the linear regression indicate whether the correlation between NIR color and veiling is statistically significant, but do not quantify the strength of the correlations: a small p-value does not necessarily imply a strong correlation. For example, a weak correlation within a large data set (N$\rightarrow \infty$) could have a very small p-value because the large number of samples drives down the statistical uncertainty. \edit1{To assess the strength of each correlation, we calculated a Spearman $\rho$ coefficient for each sample of binaries and single stars. The $\rho$ values for single and binary stars in each color are listed in Table \ref{tab:color_params}. For single stars, H-K vs. $r_{7510}$ is the strongest correlation, with $\rho_{S, H-K} = 0.66$. For binary stars, the strongest correlation is K-W1, with $\rho_{B, K-W1} = 0.68$.}

To illustrate the differences in linear regression best-fit slope between each color, and the differences between single and binary stars, Figure \ref{fig:color_slopes} shows the linear regression best-fit slope as a function of color for J-H, H-K, K-W1, and W1-W2. There are no statistically significant differences between the single and binary star best-fit slopes in the sample including $r_{7510} = 0$ points. In the regressions performed while excluding $r_{7510} = 0$ points, the difference between the single and binary slopes is only outside the mutual uncertainties in W1-W2. On average, the binary star slope is (83 $\pm$ 10)\% of the single star slope when including $r_{7510} = 0$ points, and (44 $\pm$ 25)\% when excluding $r_{7510} = 0$ points. The W1-W2 slopes are shallower than those of other colors because by W1 and W2 the disk excess SED begins to flatten out (e.g., Figure \ref{fig:source_SEDs}), meaning that the dynamic range of the color excess decreases.

The binary star best-fit slopes have smaller $\tau$ coefficients than the single stars, indicating that the binary excess is less predictive of veiling than the single star excess. This is likely because there are very few high-veiling binary stars, so the binary fit is more dominated by the scatter in the low-veiling and no-veiling sources than the single star sample, which has more high-veiling sources.

\subsection{Relationships Between Veiling and NIR Excesses}
Past work on this topic has focused on color excesses instead of single-filter excesses because calculating absolute excesses requires a robust understanding of the SED of the underlying stellar photosphere, which has historically been difficult to constrain, especially outside of the optical wavelength regime. However, individual filter excesses, when accessible, provide more specific information about the absolute, rather than the relative, disk SED. Single-filter excesses trace the disk SED, and correlations between NIR absolute excess and optical veiling reflect and quantify the relationship between the accretion column and the dust disk. Individual filter excesses also remove the zero-point ambiguity of color excesses, which are relative measurements and do not account for the J band excess caused by the disk SED \citep{Cieza2005, Fischer2011}. 

After calculating the NIR color excess, we used the \citet{Pecaut2013} bolometric correction $BC_{J}$ to calculate absolute excesses in all five filters used in this analysis. Then, we calculated a linear regression comparing each excess against the optical veiling separately for the samples of single and binary stars with disks. We calculated regressions both including and excluding systems with $r_{7510} = 0$, to examine whether quiescent disk hosts significantly impacted the relationship between NIR excess and veiling. However, we found that there was not a significant difference between the slopes calculated with and without the $r_{7510} = 0$ systems, so we concentrate on the results including the veiling-free systems.

\begin{figure*}
\plotone{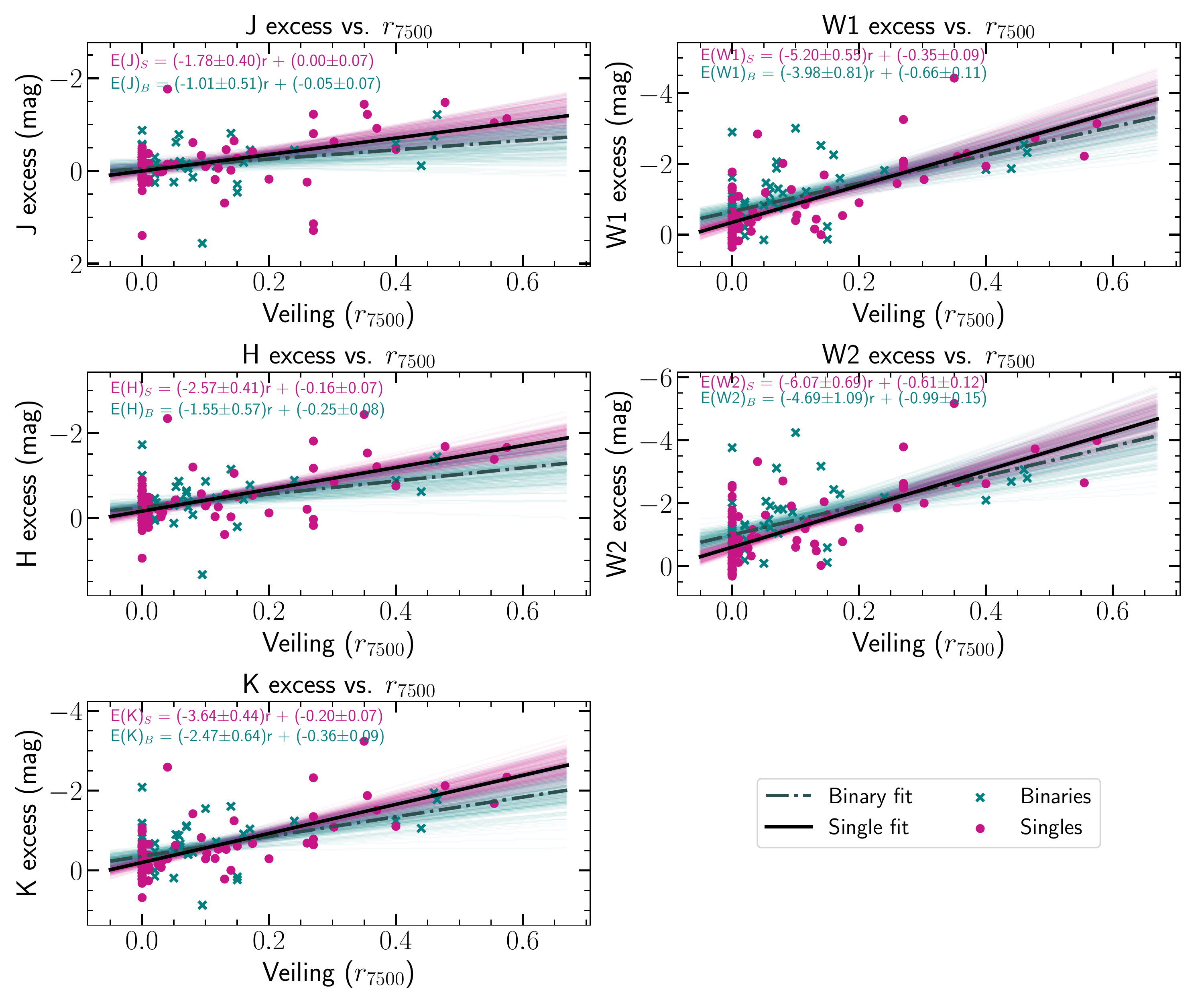}
\caption{Veiling-excess correlations for NIR filters. Binaries are shown as teal `x' markers, with the best-fit line from the linear regression overplotted as a dark blue dashed line. Single stars are shown as magenta points, with the best-fit line from a linear regression shown as the black solid line. The thin teal and magenta lines show the error on the binary and single star best-fit lines, respectively, visualized using 200 draws \edit1{from the covariance matrix returned by the linear fit}. The error bars are not plotted because they are smaller than the points. The text on each figure is the equation of the best-fit line for the single or binary stars, calculated using a linear regression. The slope of the best-fit line increases substantially moving redward across the SED.}
\label{fig:veiling_excess_corr}
\end{figure*}

The best-fit lines from each linear regression are plotted in Figure \ref{fig:veiling_excess_corr}, along with the data points used in the calculation and 200 bootstrapped samples \edit1{drawn from the covariance matrix returned from the linear fit}. The binary and single star parameters were calculated separately and are shown as teal and magenta lines, respectively. The equation of the best-fit line for the single and binary stars is also shown on each panel. We note that the excess is shown as magnitudes in excess of the photosphere, so a more negative number indicates a larger excess, and a negative slope to the linear fit indicates a positive correlation between stronger veiling and stronger NIR excess. 

\begin{deluxetable*}{cCCCCCCC}
\tablecaption{Best-fit line parameters, $\rho$ values, and P-values for each NIR magnitude \label{tab:mag_params}}
\setcounter{table}{2}
\tablecolumns{8}
\tablewidth{0pt}
\tablehead{
\colhead{Filter} & \colhead{Slope} & \colhead{Y-intercept} & \colhead{P-value} & \colhead{Spearman $\rho$} & \colhead{Slope} & \colhead{Y-intercept} & \colhead{Y-intercept}\\
\colhead{} & \colhead{$f(r_{7510})$} & \colhead{$f(r_{7510})$} & \colhead{$f(r_{7510})$} & \colhead{$f(r_{7510})$} & \colhead{(no $r=0$)} & \colhead{(no $r=0$)} & \colhead{$f(r_{7510}-\overline{r_{7510}})$\tablenotemark{a}}\\
}
\startdata
\multicolumn{8}{c}{Single Stars}\\
\hline
J & -1.78$\pm$0.40 & 0.00$\pm$0.07 & 0.001 & -0.37 & -1.82$\pm$0.62 & 0.01$\pm$0.14 & -0.13$\pm$0.04\\
H & -2.57$\pm$0.41 & -0.16$\pm$0.07 & 0.0001 & -0.45 & -2.55$\pm$0.61 & -0.17$\pm$0.14 & -0.35$\pm$0.04\\
K & -3.64$\pm$0.44 & -0.20$\pm$0.07 & $<0.0001$ & -0.60 & -3.55$\pm$0.65 & -0.23$\pm$0.15 & -0.47$\pm$0.04\\
W1 & -5.20$\pm$0.55 & -0.35$\pm$0.09 & $<0.0001$ & -0.67 & -4.85$\pm$0.77 & -0.46$\pm$0.18 & -0.72$\pm$0.04\\
W2 & -6.07$\pm$0.69 & -0.61$\pm$0.12 & $<0.0001$ & -0.66 & -5.56$\pm$0.88 & -0.77$\pm$0.20 & -1.05$\pm$0.04\\
\hline
\multicolumn{8}{c}{Binary Stars}\\
\hline
J & -1.01$\pm$0.51 & -0.05$\pm$0.07 & $>0.05$ & -0.10 & -1.63$\pm$0.75 & 0.11$\pm$0.14 & -0.13$\pm$0.06\\
H & -1.55$\pm$0.57 & -0.25$\pm$0.08 & $>0.05$ & -0.26 & -2.15$\pm$0.77 & -0.09$\pm$0.14 & -0.36$\pm$0.06\\
K & -2.47$\pm$0.64 & -0.36$\pm$0.09 & 0.01 & -0.33 & -2.83$\pm$0.82 & -0.27$\pm$0.15 & -0.54$\pm$0.06\\
W1 & -3.98$\pm$0.81 & -0.66$\pm$0.11 & $<0.0001$ & -0.60 & -3.23$\pm$0.98 & -0.85$\pm$0.18 & -0.94$\pm$0.06\\
W2 & -4.69$\pm$1.09 & -0.99$\pm$0.15 & $<0.0001$ & -0.60 & -3.27$\pm$1.27 & -1.36$\pm$0.23 & -1.33$\pm$0.06\\
\enddata
\tablenotetext{a}{The y-intercept of the line calculated using a regression with the x-values centered at the mean of the veiling values, which is $\overline{r_{7510}} = 0.072$..}
\end{deluxetable*}

Table \ref{tab:mag_params} shows the relevant values from the linear fit, including the y-intercept calculated using a linear regression with the data points centered at the origin \edit1{and the linear parameters for a fit excluding the $r_{7510} = 0$ points}, as discussed in Section 4.1. From the linear regression we find that all filters have significant p-values (p$<$0.05) for both single and binary stars, indicating that the data are correlated, except for the J \edit1{and H} band binary relations. To assess the strength of the correlation between veiling and individual filter excess \edit1{we calculated the Spearman $\rho$ coefficient, also listed for each fit in Table \ref{tab:mag_params}. Because the data are in magnitudes, a more negative $\rho$ corresponds to a stronger positive correlation between veiling and NIR excess. We find that the binary star fits are systematically more weakly correlated than single stars, and that the W1 filter shows the strongest correlation between veiling and NIR excess, with Spearman $\rho$ values of -0.67 and -0.60 for single and binary stars, respectively.}

\begin{figure*}
    \plottwo{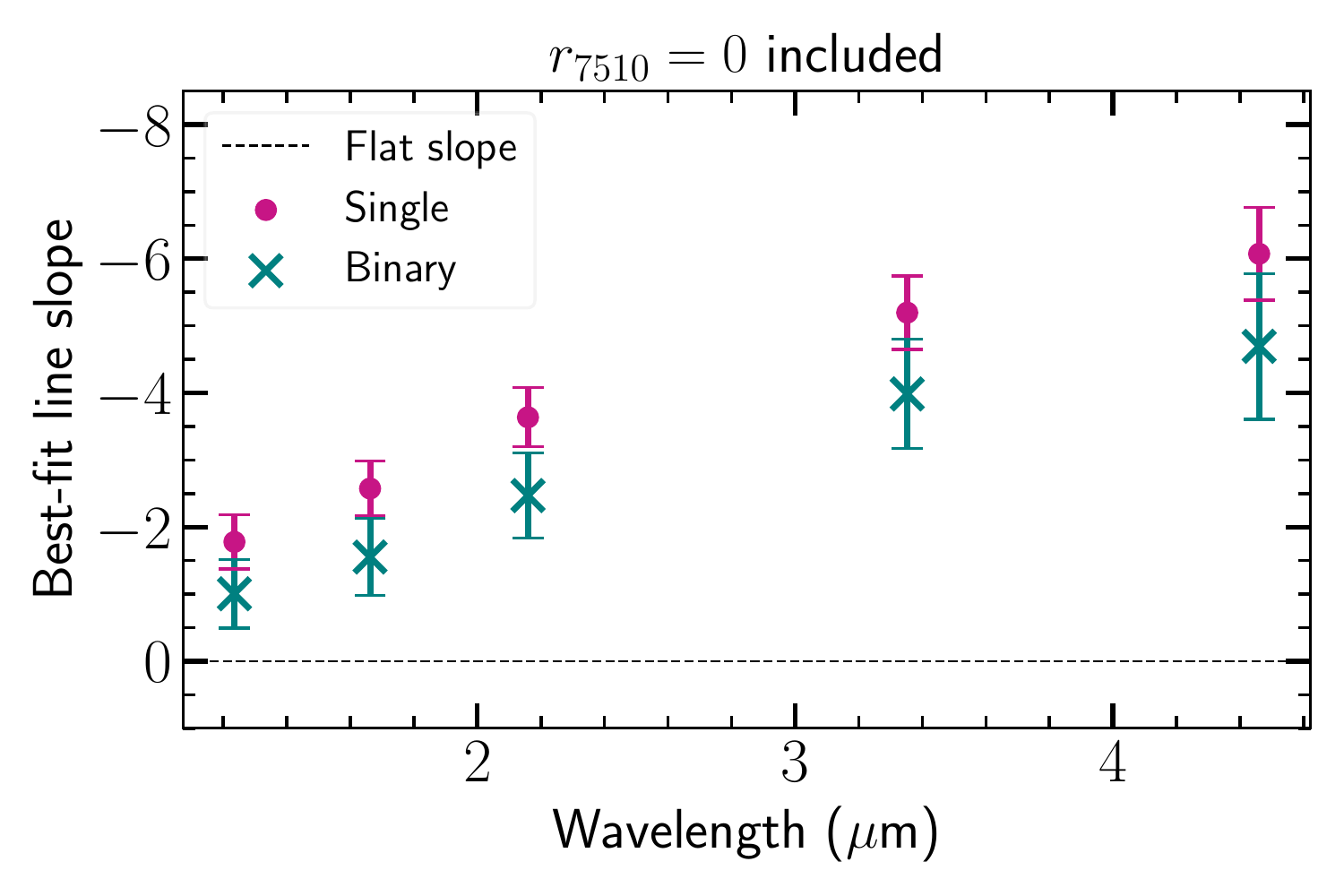}{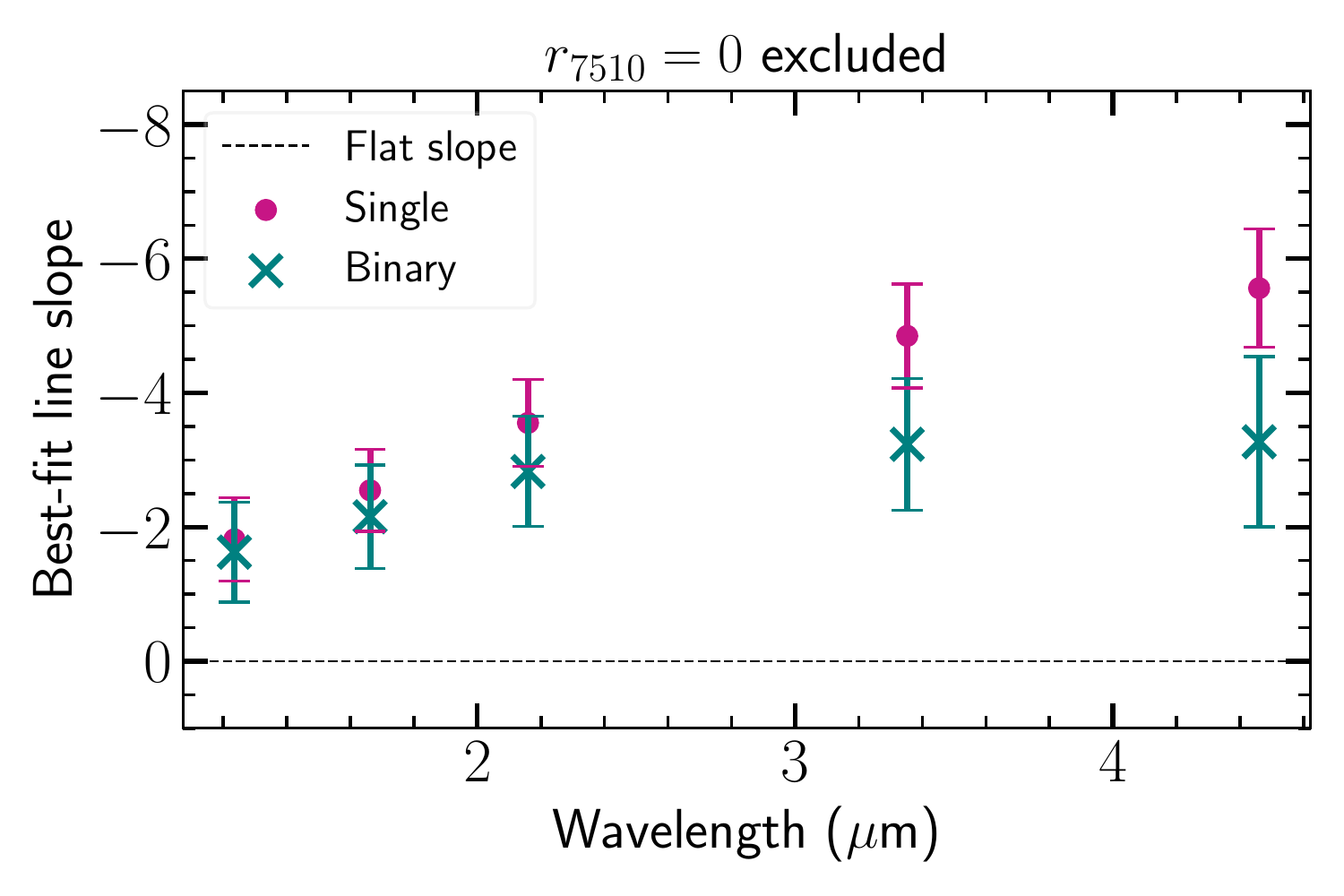}
    \caption{The slopes of the best fit line to the relationship between absolute NIR excess and optical veiling for binary and single stars (teal and magenta, respectively), where the error bars are the RMS error on the slope with (left) and without (right) $r_{7510} = 0$ systems included. \textbf{Left:} When including $r_{7510} = 0$ points in the linear regression, there are no statistically significant differences between the single and binary slopes across all wavelengths. \textbf{Right:} After excluding the $r_{7510} = 0$ points, the differences between the single and binary sample slopes remain statistically insignificant except in W2.}
    \label{fig:mag_slopes}
\end{figure*}

Figure \ref{fig:mag_slopes} shows the slopes of the binary and single star linear regressions as a function of wavelength both including (left) and excluding (right) $r_{7510} = 0$ systems. As the circumstellar disk becomes the dominant source of flux in the SED, the dynamic range of the excess increases, causing the slope of the best-fit line to increase across the SED. When the $r_{7510} = 0$ points are included, although some of the binary and single star slopes have overlapping error bars, on average the binary star slope is slightly shallower than the single star slope ($m_{binary} \approx 0.7 m_{single}$, although the relative difference is clearly not constant; Figure \ref{fig:mag_slopes}). When the $r_{7510} = 0$ points are excluded from the regression, the slopes are slightly shallower, and the differences in slope between binary and single stars become more significant toward the red rather than remaining relatively constant as in the case including the $r_{7510} = 0$ points. However, the single and binary stars only disagree by more than the mutual uncertainty in the reddest filter (W2).

\begin{figure*}
    \plottwo{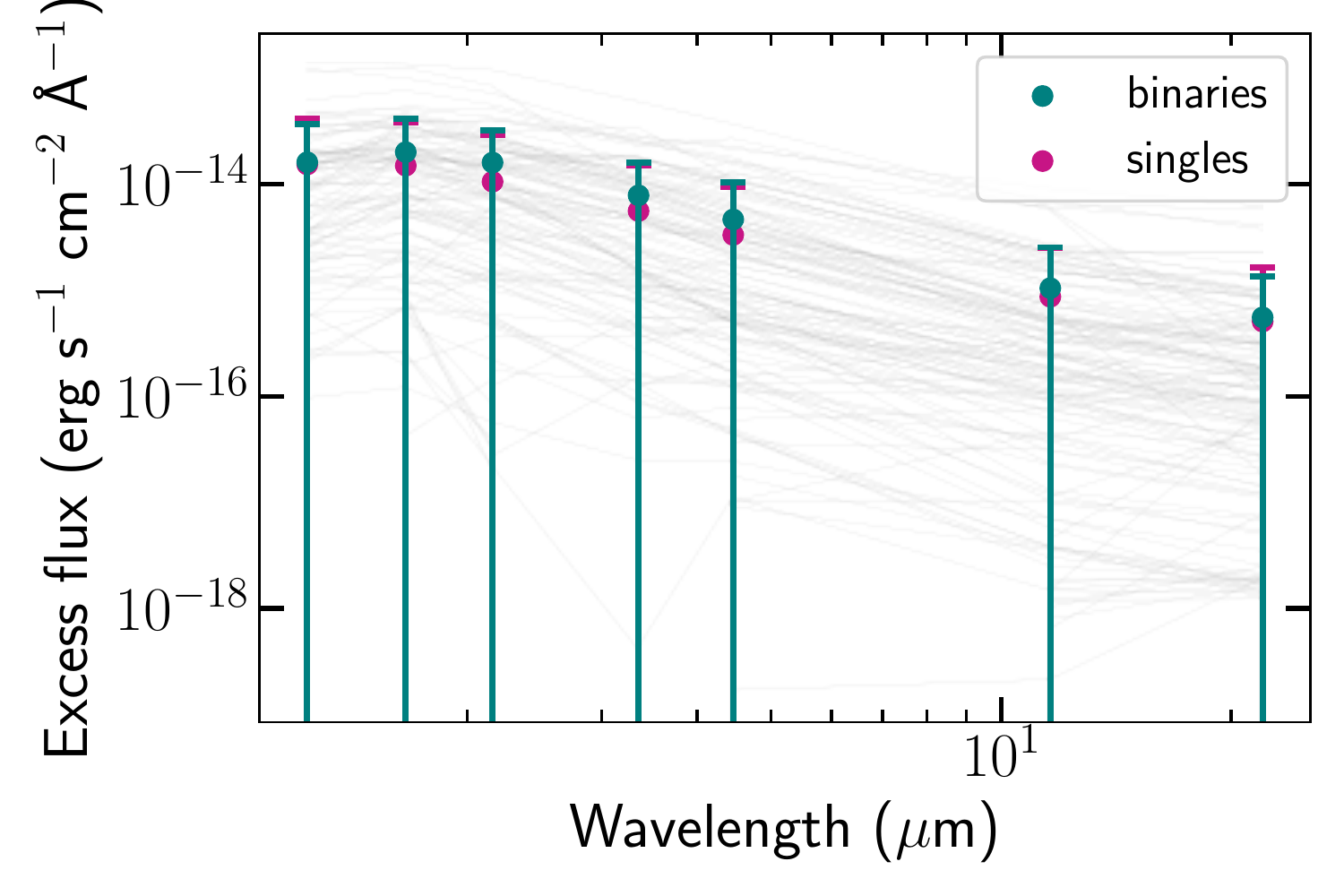}{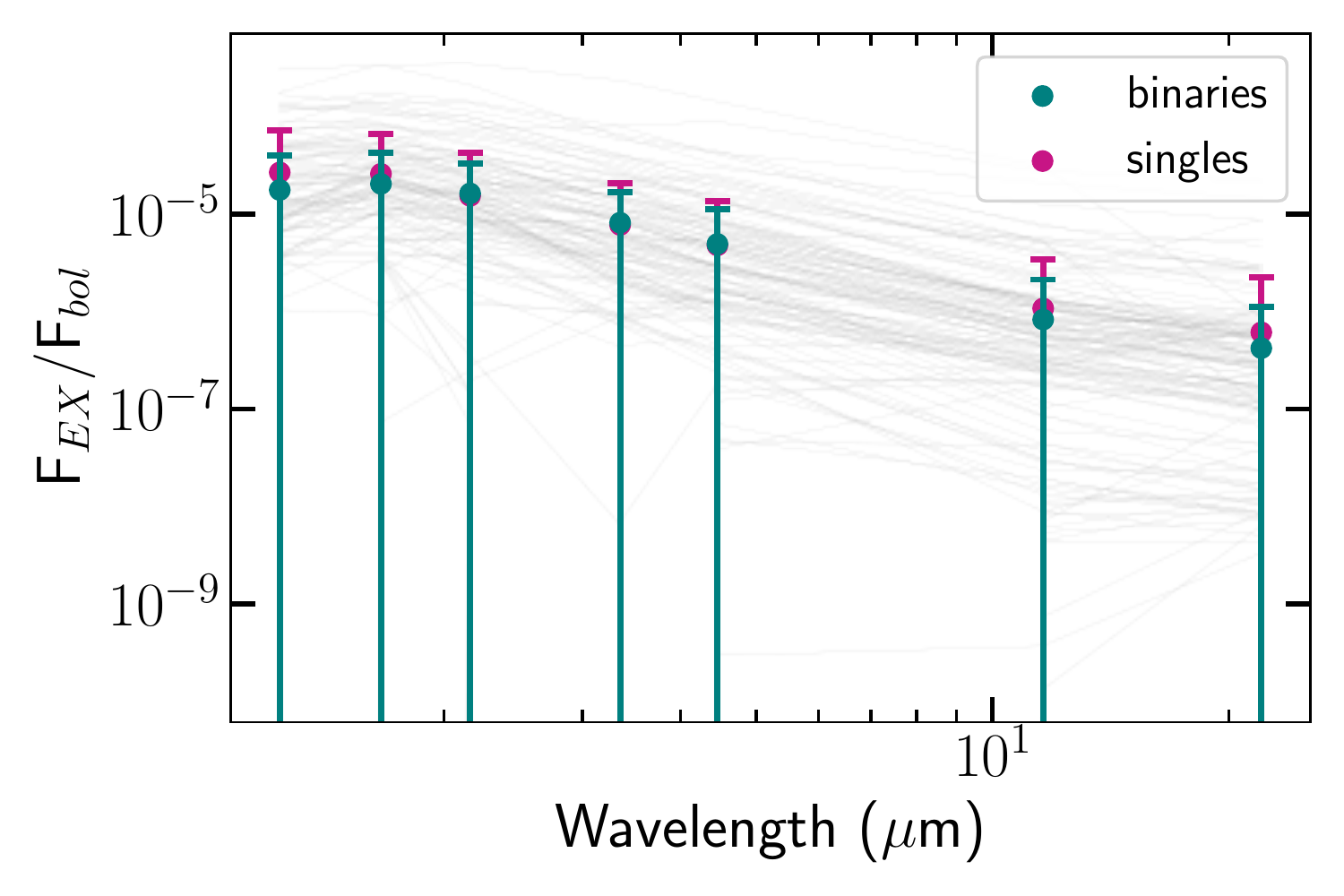}
    \caption{Left: The SED points of the mean circumstellar disk for single (magenta) and binary (teal) stars in our sample, produced by converting each filter excess into a flux excess. The error bars show the RMS scatter of the sample, and the gray lines underlaid show each individual system's SED. There is large rms scatter among all the systems, but the mean binary disk flux is slightly larger than the single disk flux. Right: The same figure as on the left, but with all filter fluxes normalized by each system's bolometric flux. Because increased F$_{bol}$ should cause a larger dust truncation radius at the same temperature, increasing the amount of disk emission, this normalization attempts to remove that dependence. Across the middle of the SED binary disks remain brighter than single stars, but at each end of the SED binaries are fainter than singles.}
    \label{fig:flux_SED}
\end{figure*}

To better demonstrate the structure of the disk SED, we converted the magnitude excesses for each system into flux excesses, and plotted the resulting mean and standard deviation of each filter, shown in the left panel of Figure \ref{fig:flux_SED}. The gray lines underlaid on the figure show the SEDs for individual systems. The error bars do not show the error on the measurement, but rather show the RMS spread in the distribution of samples. The SED peak falls at the H band for both single and binary stars, which is the location of the peak emission of a blackbody at a temperature of $T_{eff} \approx 1800$K, which is consistent with the expected dust sublimation temperature of $\sim 1500-2000$K. The mean binary star disk SED is typically slightly brighter (F$_{binary} \approx 1.2$ F$_{single}$) than the single star SED, except in J band, where F$_{binary} \approx 0.8$ F$_{single}$. However, a Kolmogorov-Smirnov statistic test \citep[K-S test;][]{Kolmogorov1933, Smirnov1939, Massey1951, Hodges1958} finds that the p-value for most filters is greater than 0.05 (the standard cutoff for statistical significance), indicating that there is no statistically significant difference in the disk flux distributions between single and binary stars. 

The right panel of Figure \ref{fig:flux_SED} shows the same data as the left panel, but normalized by the bolometric flux of each host star \edit1{(calculated using the rescaled $L_{bol}$ and \textit{Gaia} distance)}. The emitting area of the inner edge of the disk should increase with stellar flux, because more radiation will move the dust sublimation radius to larger distances, so normalizing by F$_{bol}$ attempts to correct for system-to-system variations in brightness and emitting area, allowing more direct comparison. In the normalized case, binary disks remain brighter than single disks in $KW1W2$ by $\sim 25\%$ but are fainter in $JH$ and $W3W4$ by $\sim 20\%$. From K-S tests, the normalized disk flux distributions are still statistically indistinguishable. 

\section{Discussion} \label{sec:disc}
To understand the relationship between optical and infrared excess in pre-main sequence stars, we have assembled a large, homogeneously characterized sample of Taurus members using the measurements of HH14, calculated the infrared color and single-filter excesses for single and close binary stars using 2MASS and WISE observations ($JHK_{s}W1W2$ filters and J-H, H-K, K-W1, and W1-W2 colors) and investigated possible relations between the optical veiling and the NIR excess in both individual filters and NIR colors. 

We found that NIR excess (JHKW1W2 filters, or $1.1 < \lambda < 4.5 \mu$m) is correlated with optical veiling across all bands and colors. Our findings that NIR and optical excesses are related are consistent with previous work \citep[e.g.,][]{Hartigan1990, Hartigan1991, Kenyon1995, Cieza2005}, and continued to explore the bluer region of the NIR (J and H bands, or $\lambda <2.2 \mu$m), where the origins of the NIR emission are still not known \citep{Cieza2005, Fischer2011, McClure2013}. We found that H-K is the NIR color that is most strongly correlated with veiling. This is consistent with a dust sublimation origin for some of the NIR excess, because H-K is the color associated with of the peak of a blackbody at the dust sublimation temperature, and thus should have the largest sensitivity to emission arising from the inner edge of the dust disk. We further discuss the relationship between veiling and NIR excess in Section \ref{sec:nir disc}.

We found that the W1 single-filter excess is most strongly correlated with veiling. This is because the relation is nearly as steep as the W2-veiling relationship, but has significantly less scatter, likely because W1 is closer to the disk SED peak and so is brighter than W2. We found that many of the systems that meet the \citet{Rebull2010} disk host criteria have no veiling measured, and we discuss these systems in Section \ref{sec:noveil}. We found that after converting to fluxes, there is no significant difference between single and binary star disk SEDs. We discuss the binary subsample in Section \ref{sec:binary_disc}. We also found that in a sample of active accretors (systems with r $>$ 0) the binary relation is significantly shallower than the single star relation at long wavelengths, indicating that in actively accreting binaries the inner disk rim emission is not as strongly correlated with accretion as it is for single stars or when also including quiescent binary stars.

\subsection{The Relationship Between NIR Excesses and Optical Veiling}\label{sec:nir disc}
The observed correlation between NIR excess and optical veiling means that larger disk flux is either directly or indirectly related to a higher stellar accretion rate. The relation between larger excess and higher veiling indicates that sources with more active accretion also produce more NIR emission, both in relative (i.e., colors) and absolute (i.e., disk flux) measures. Because colors indicate the flux ratio between two SED points, the correlation between NIR colors and optical excess corresponds to a brighter and/or steeper SED. Physically, this relates to the temperature and luminosity of the inner disk. Similarly, disk fluxes and absolute excesses correspond to the total emitting area of the inner disk, because the dust sublimation temperature is roughly constant, so an increase in disk flux at a given wavelength can only be produced by a larger emitting area. 

The two dimensions over which to increase the disk emission area are the dust disk inner radius and the scale height of the inner disk. The dust disk inner radius is set by the total flux emitted from the system, of which the accretion luminosity is a small fraction ($L_{acc} \sim 0.1 L_{\star}$; \citealt{Alcala2014, Alcala2017, Robinson2019}), and stellar optical variability does not appear to induce NIR variability \citep{Flaherty2012}, so the NIR variations do not appear to be caused by an increase in the dust disk inner radius. Thus, \citet{Flaherty2012} concluded that the NIR variability they observed was caused by increases in the height of the inner rim of the dust disk, which is inflated because of increased internal heating \citep{Dullemond2001, Isella2005}. Possible mechanisms \citet{Flaherty2012} suggested for increasing scale heights include changes to the stellar magnetic field, X-ray flares, and companions perturbing the disk.

\citet{Espaillat2019} did not observe a correlation between accretion rate and X-ray flux, but did observe a correlation between the accretion rate $\dot{M}$ and the MIR variability of GM Aur. Because of this correlation, \citet{Espaillat2019} suggest that increasing surface density of optically thin dust in the inner disk increases both disk emission and accretion rates by inducing additional mass loading onto the stellar magnetosphere. However, the timescale of this correlation was observed to be on the order of days to weeks, and therefore may not fully explain the correlation we observe between accretion rate and disk emitting area in asynchronous observations. The persistence of this correlation suggests that there is a relatively steady-state process causing both the NIR excess and the average veiling, even though they are variable on short timescales. One possibility is that more massive dust disks have larger inner walls due to increased internal pressure, while disk mass is correlated with accretion rate \citep{Manara2016mdot}, causing increased veiling. Synchronous observations with longer time baselines will help clarify the origin of the asynchronous correlation between veiling and NIR excess. Performing a similar analysis of additional star forming regions, such as the well-defined sample of disk hosts in Upper Scorpius \citep{Luhman2012,Esplin2018} or the Lupus sample of \citet{Alcala2017}, could explore whether these relationships and statistics are shared between regions of different ages and with different environments.

\subsection{NIR Excesses in Veiling-Free Stars}\label{sec:noveil}
\begin{figure*}
    \plotone{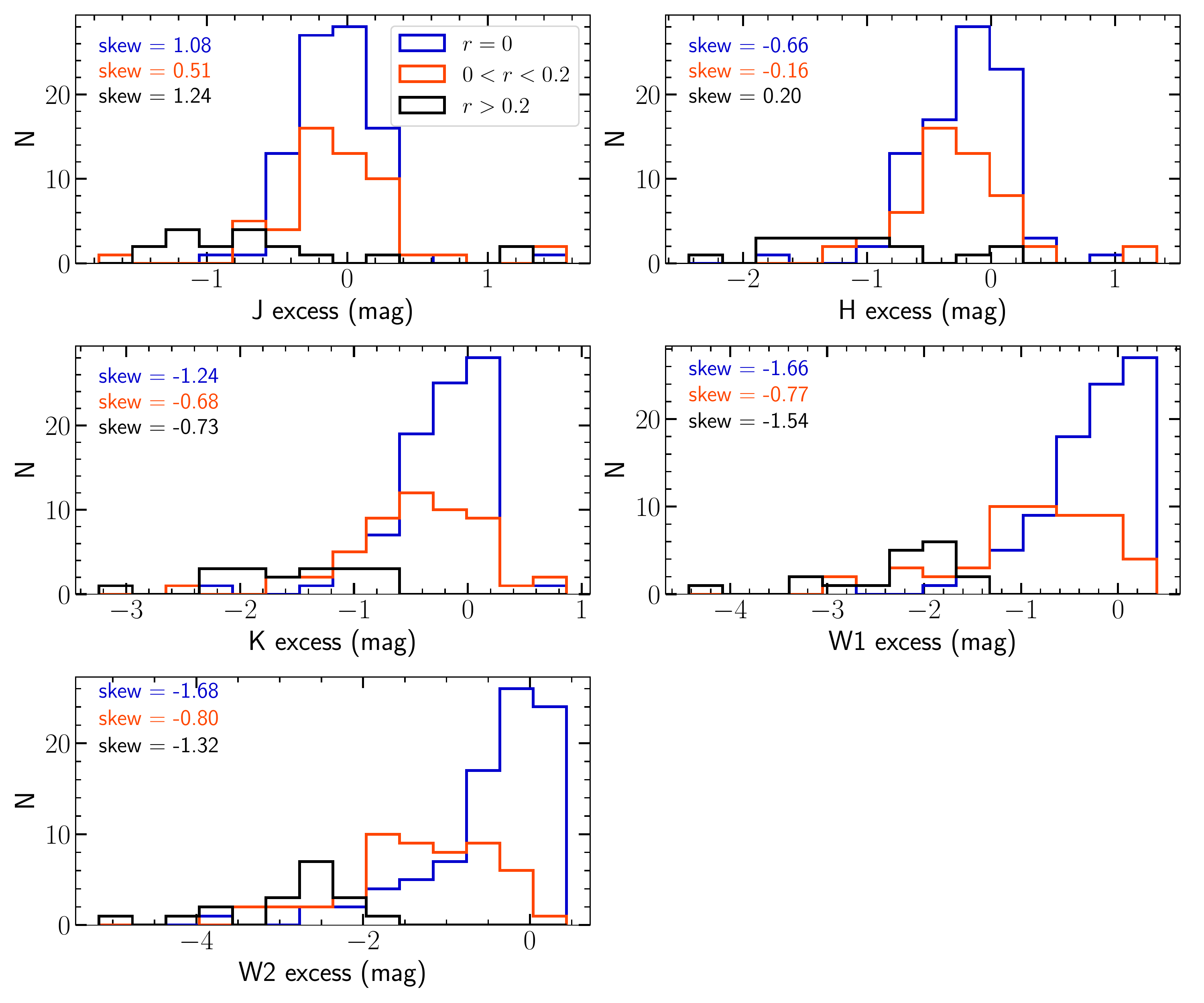}
    \caption{Histograms of the single-filter NIR excess distribution for disk-hosting sources with veiling values of zero (blue), $0 < r \leq 0.2$ (orange), and $r > 0.2$ (black). \edit1{The text on each figure shows the skew value for each histogram, color-coded appropriately.} The zero-veiling sources are typically centered around an excess value of zero for all filters, while the higher-veiling sources have progressively larger average excess. The asymmetry \edit1{(skewness)} of all histograms increases with increasing wavelength, showing that the veiling-free systems still have excess flux from a disk. We note again here that in absolute NIR excesses, a negative value indicates a flux excess.}
    \label{fig:noveil_hist}
\end{figure*}

We have found that NIR excess and veiling are correlated for stars that meet the \citet{Rebull2010} criteria for being a disk host. However, there were many disk host stars in our sample with little or no veiling measured, and that there was significant scatter in the excesses measured for the zero-veiling sources, indicating that a source with no veiling could still have a large amount of NIR excess.

We measured NIR excesses consistent with the presence of a disk in 133 of our 163 systems. Of those 133 systems, 64 ($\sim 50\%$) had veiling measurements of $r_{7510} = 0$. The systems with no veiling are coeval with systems that show both veiling and a disk ($\tau_{r = 0} = 3.14$Myr; $\tau_{r \neq 0} = 3.22$Myr), indicating that these systems are not necessarily older than their veiled counterparts, although we caution that age is not necessarily an indicator of evolutionary stage and age can be difficult to infer for young stars.

\citet{Natta2006} measured the accretion properties of a sample of Ophiuchus PMS stars using hydrogen emission lines in the NIR (Pa$\beta$ and Br$\gamma$), and similarly found that $\sim$50\% of their targets exhibited signs of hosting a disk but did not have measureable accretion. The optical continuum excess/veiling measurement and the NIR emission line measurement measure the same relative number of systems with significant levels of accretion and systems with disks. This suggests that both methods may have the same level of sensitivity to accretion. The strength of optical veiling and the equivalent width of NIR emission lines are correlated \citep[e.g.,][]{Muzerolle1998, Calvet2004, Alcala2014, Alcala2017}, and have similar origins. The optical/UV emission arises from the accretion shock on the stellar surface, while the NIR emission lines are produced in the accretion column \citep{Calvet2004}. Similar statistics in the number of systems lacking each of these properties while still hosting disks implies that there is little to no accretion occurring within the magnetospheric truncation radius (i.e., neither in the accretion column nor in the accretion shock) in these systems, which is consistent with the physical origins of the NIR emission lines and optical veiling. However, no measureable accretion occurring at one time does not necessarily mean that accretion has permanently ceased.

The discrepancy between the numbers of systems with disks and systems with measured veiling may be a result of some disks beginning to evolve, causing accretion to fall below detectable levels, or could be caused by the intrinsic variability of accretion causing large changes in veiling. For example, \citet{Venuti2014} observed veiling variation on the order of 0.5 dex on timescales of weeks. More direct indicators of accretion also vary rapidly: accretion rates derived from H$\alpha$ can vary on timescales of hours to days; \citep{Mundt1982, Johns1995, Costigan2012, Sousa2016, Sousa2021}. It seems most likely that the observed high fraction of veiling-free disk hosts is caused by a combination of these factors.

As circumstellar disks evolve, the slope of the MIR SED increases more rapidly with wavelength as dust grains settle and thus intercept less stellar radiation \citep{Dullemond2004} and a central, more transparent cavity clears because of grain growth and possibly planet/planetesimal formation in the inner disk \citep{Dullemond2005}. The combination of these factors means that a more evolved circumstellar disk should continue emitting at longer wavelengths even after the inner disk has begun to dissipate.

Figure \ref{fig:noveil_hist} shows histograms of the magnitudes of excess in each NIR filter used in our analysis for systems with $r_{7510} = 0$, $0 < r_{7510} \leq 0.2$, and $r_{7510} > 0.2$. The majority of systems without measured veiling have little to no NIR excess, with most of the distributions centered around 0 excess. Moving through the NIR to longer wavelengths, the distributions become progressively less symmetric \edit1{(e.g., when assessed using a skew test, we found the skewness increases with increasing wavelength, as demonstrated by the values listed on Figure \ref{fig:noveil_hist})}, indicating that systems without veiling have increasing amounts of excess emission at progressively longer wavelengths, as would be expected from more evolved disks. This behavior is in contrast to the higher-veiling systems, which show a similar but much less dramatic asymmetry toward longer wavelengths that begins earlier than in the $r_{7510} = 0$ case. The disk-clearing process is rapid \citep[e.g.,][]{Simon1995, Wolk1996, Andrews2005, Luhman2010, Alexander2014}, leading to small numbers of observed transitional disks \citep[e.g.,][]{Kenyon1995, Luhman2010}, meaning that we would not expect nearly 50\% of our sample to be transitional disks. However, it is unclear if there is another explanation for the asymmetric NIR excess distribution for the veiling-free sample.

In addition to disk evolution impacting accretion and disk properties, accretion itself is not a steady-state process, and stochastic variations in accretion cause photometric and veiling variability. For example, HH14 measured veiling values that changed by up to a factor of three on timescales of days, while \citet{Venuti2014} found that the UV/optical excess rarely varied by more than 0.5 dex on timescales of a week. Similarly, \citet{Fang2020} found that large changes in optical veiling are rare but that variability is common. Changes in accretion, and thus in veiling, may also significantly affect the number of systems with measured veiling relative to the number of systems with disks. Without time-domain observations of our sample of nominally veiling-free sources it is impossible to determine the relative contributions of variability or evolutionary stage to the observed lack of veiling in nearly half of our sample. 

\subsection{The Impact of Photometric Variability on NIR Excesses}
Young stars are photometrically variable, which compounds with veiling variability to introduce scatter to our asynchronous comparisons between NIR colors and optical veiling. However, although young stars are variable, the amplitude of photometric variability is almost always $\Delta \text{mag} < 1$mag, and typically closer to $\Delta \text{mag} < 0.2$ mag \citep{Bouvier1995, Carpenter2001, Herbst2002, Cody2013, Sousa2016} in both the optical and the NIR. Some young stars, such as RW Aur A+B, AA Tau, and V409 Tau in Taurus, as well as the UXor and EXor classes of variable stars, are known to have much greater variability \citep[up to or greater than $\sim 4$ mag;][]{McLaughlin1946, Herbig1950, Bouvier1999, Beck2001, Herbig2008, Bouvier2013, Rodriguez2013, Holoien2014, Rodriguez2015, Rodriguez2016, Lamzin2017}; however, such systems are rare \citep{Herbst1994, Grankin2007, Herbig2008} and so we do not expect them to significantly impact our results.

\subsection{The Effect of Multiplicity on NIR and Optical Excess}
\label{sec:binary_disc}
We have combined a large sample of homogeneously characterized Taurus members with adaptive optics multiplicity surveys. We did not modify our sample based on any additional binary properties (e.g., disk properties or separation), meaning that the binary sample includes systems with a wide variety of configurations and disk properties. For example, CoKu Tau/4 is a very close binary with a cleared central cavity created by the binary tidal forces, creating an SED that resembles a transition disk \citep{Ireland2008}. Another example of unusual binary configuration is DF Tau, and likely others, where only one star in the binary has a disk \citep{Allen2017}.

Our compilation of Taurus binaries allowed us to quantify the differences in inner disk and accretion properties between single and close binary stars for the first time, which are expected to occur because of the impact of the secondary companion on the evolution of the disk(s) of binary systems. We found that binary star NIR excesses are not as strongly correlated with veiling as single star NIR excesses, and that the differences between single stars and binary stars are most significant in significantly accreting systems where $r_{7510} > 0$. In this subsection we explore the implications of the differences and similarities between the single and binary star samples.

\begin{figure*}
\gridline{\fig{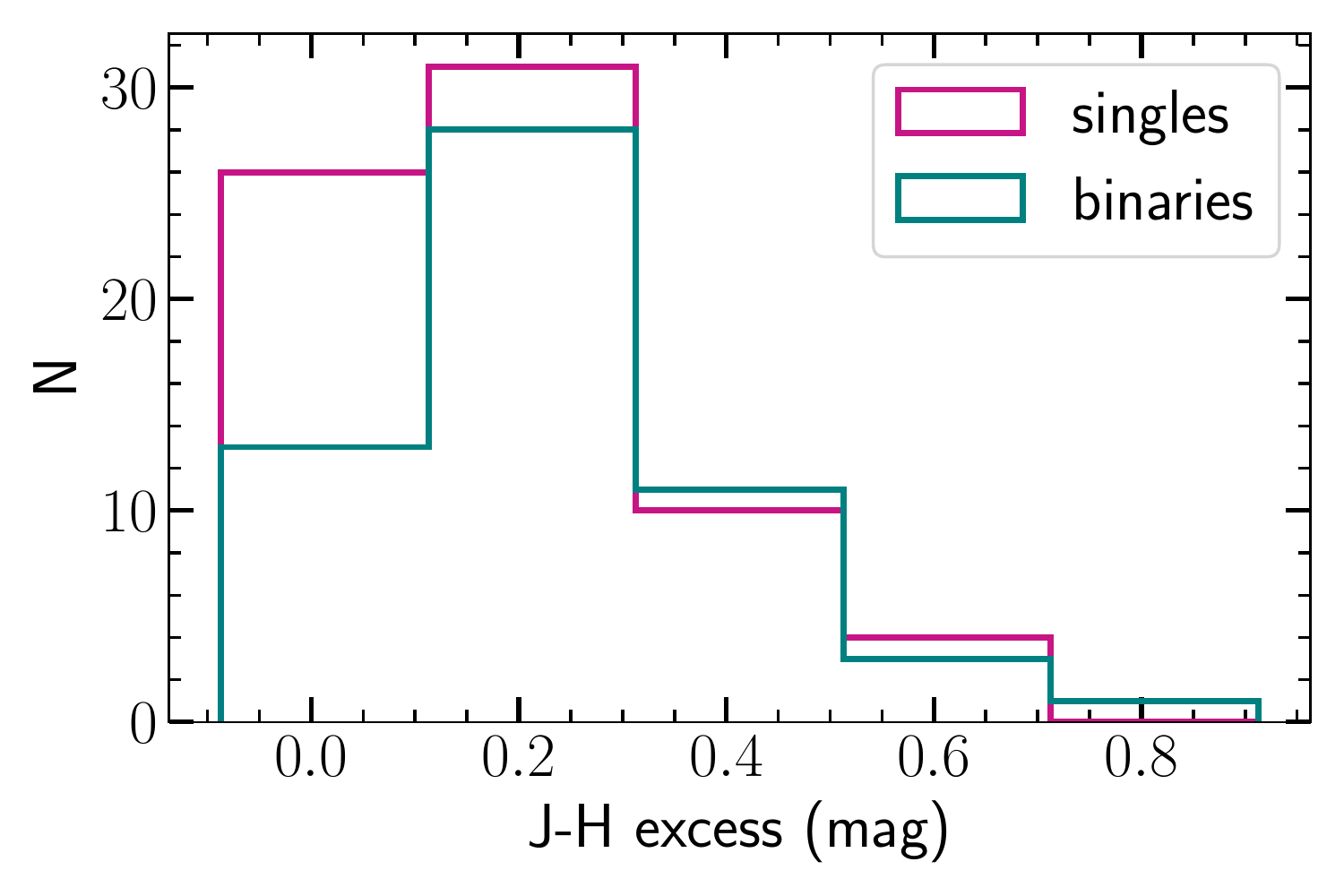}{0.45\linewidth}{}
    \fig{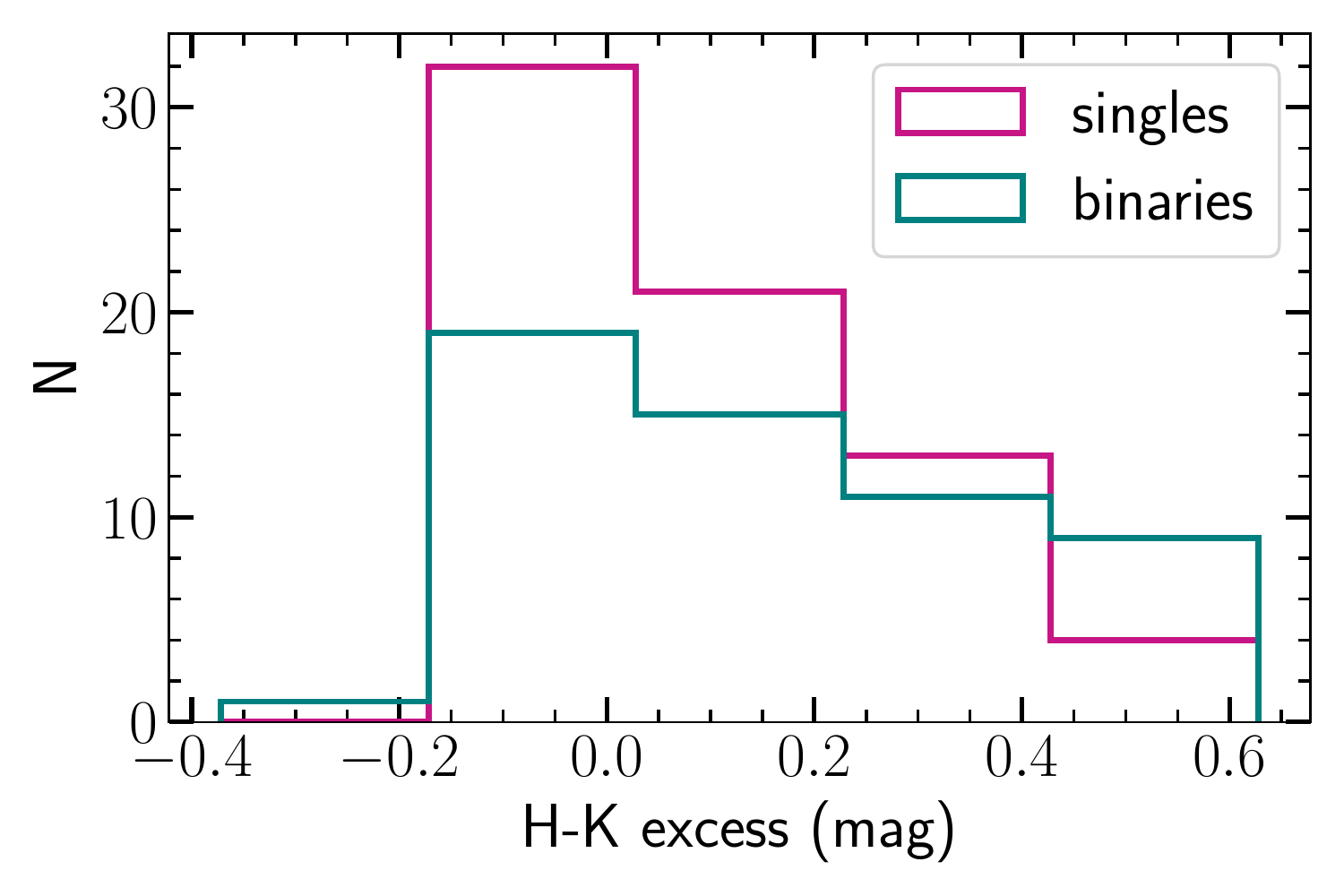}{0.45\linewidth}{}
	}
\gridline{\fig{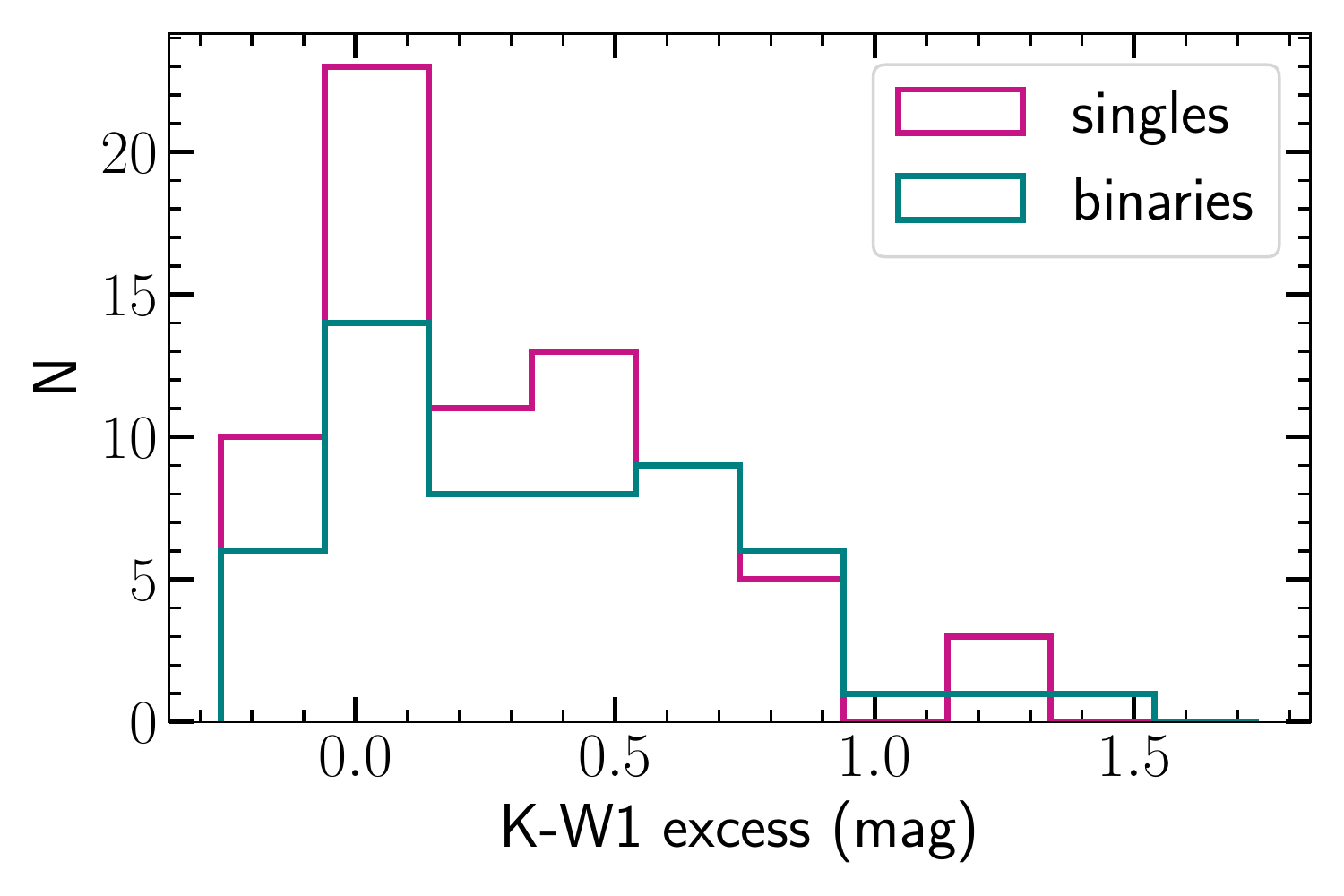}{0.45\linewidth}{}
    \fig{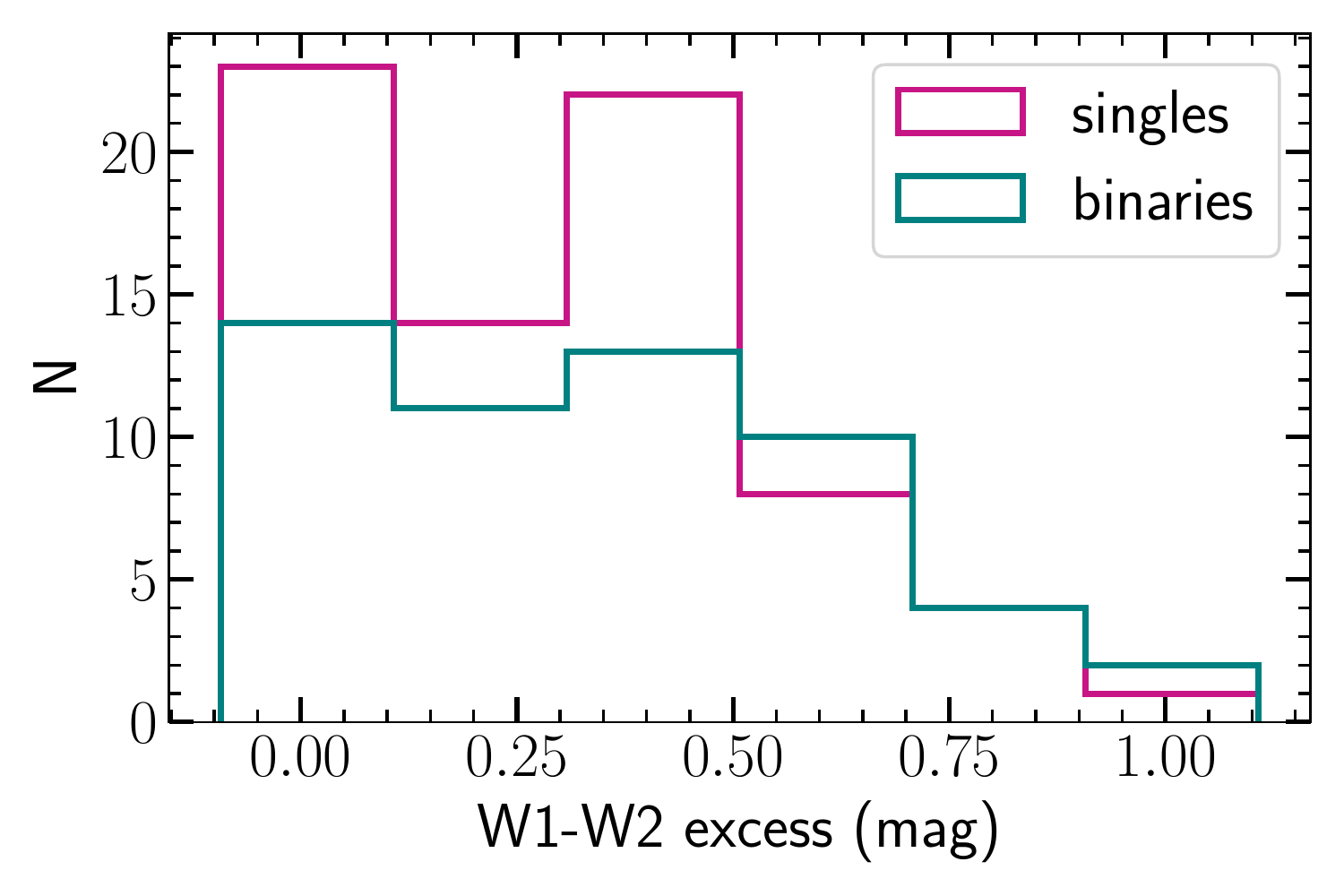}{0.45\linewidth}{}
	}
\gridline{
    \fig{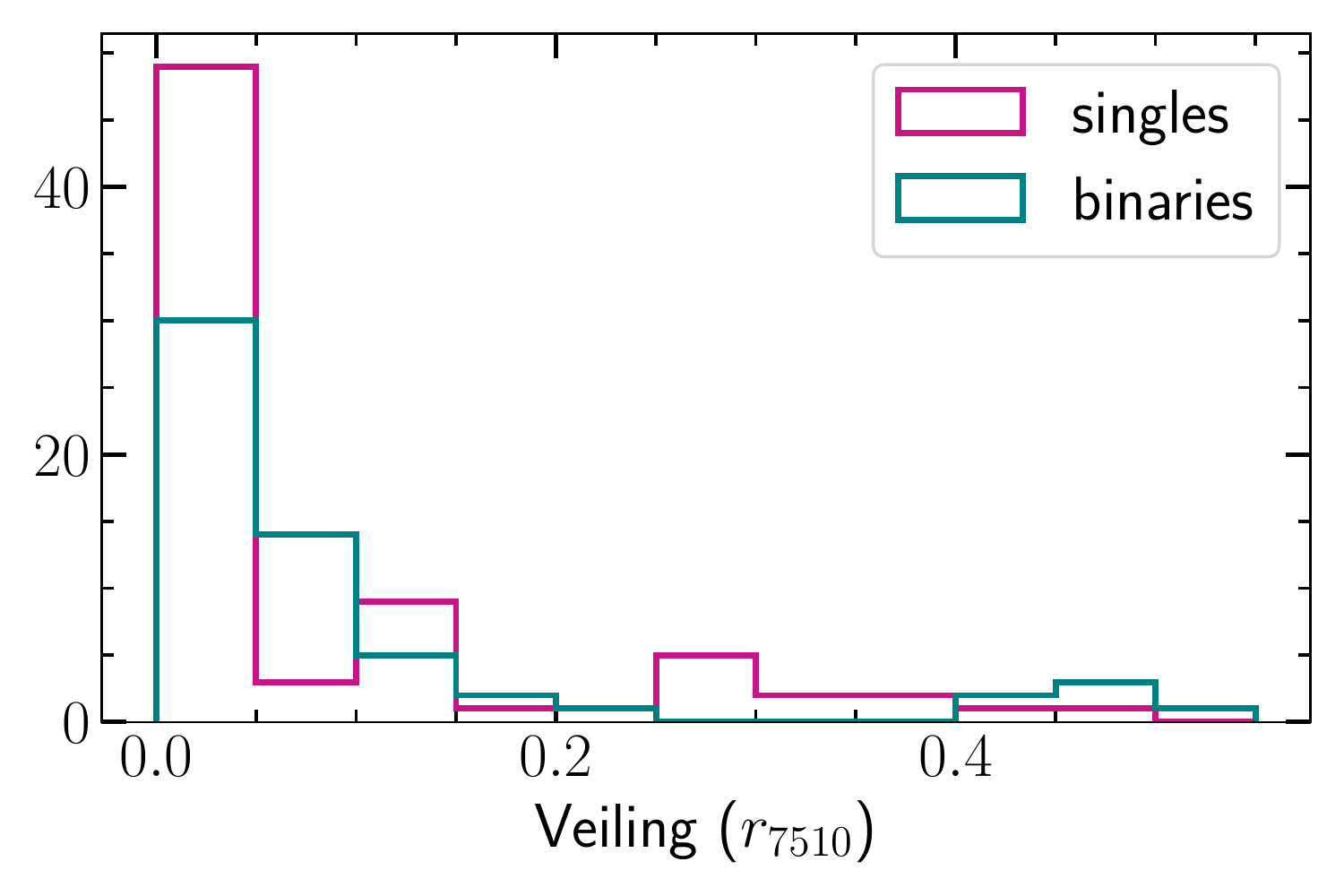}{0.45\linewidth}{}
	}
\caption{Top two rows: Histograms of color excesses for single and binary stars. The binary and single star excess distributions pass a K-S test, indicating that the distributions are drawn from the same underlying population. Bottom row: Histogram of veiling values for single and binary stars. The two samples also pass a K-S test, indicating that the two samples are drawn from the same underlying distribution.}
\label{fig:bin_single_compare}
\end{figure*}

Because we measured the veiling-color excess relation separately for single and binary stars and found different slopes for each fit, we wished to explore whether the individual data sets used to construct the relation were different between single and binary stars. Figure \ref{fig:bin_single_compare} shows histograms of the color excesses (top two rows) and veiling (bottom row) of binaries and single stars separately. We found by K-S tests (which are sensitive to the center of a distribution) and Anderson-Darling tests (which are sensitive to the edges of a distribution) that the single and binary star samples had identical color excesses and veiling distributions (e.g., p$_{K-S}$ = 0.35 and 0.2 for H-K excess and veiling, respectively; p$_{A-D} > 0.25$ for all parameters). This suggests that the binary and single star disk SED distributions are statistically indistinguishable, meaning that any differences between the single and binary stars are likely not related to differences in the input distributions.

We found that the binary star veiling-color excess and veiling-filter excess Pearson r values (Tables \ref{tab:color_params} and \ref{tab:mag_params}) are smaller than those of the single stars, indicating that NIR excess is more weakly correlated with veiling for binary stars. This may be caused by a physical property of binaries, but could also be a sign that the dynamic range of binary excesses is smaller than that of single stars: a larger dynamic range will increase the apparent level of covariance between two parameters.

\begin{figure}
    \plotone{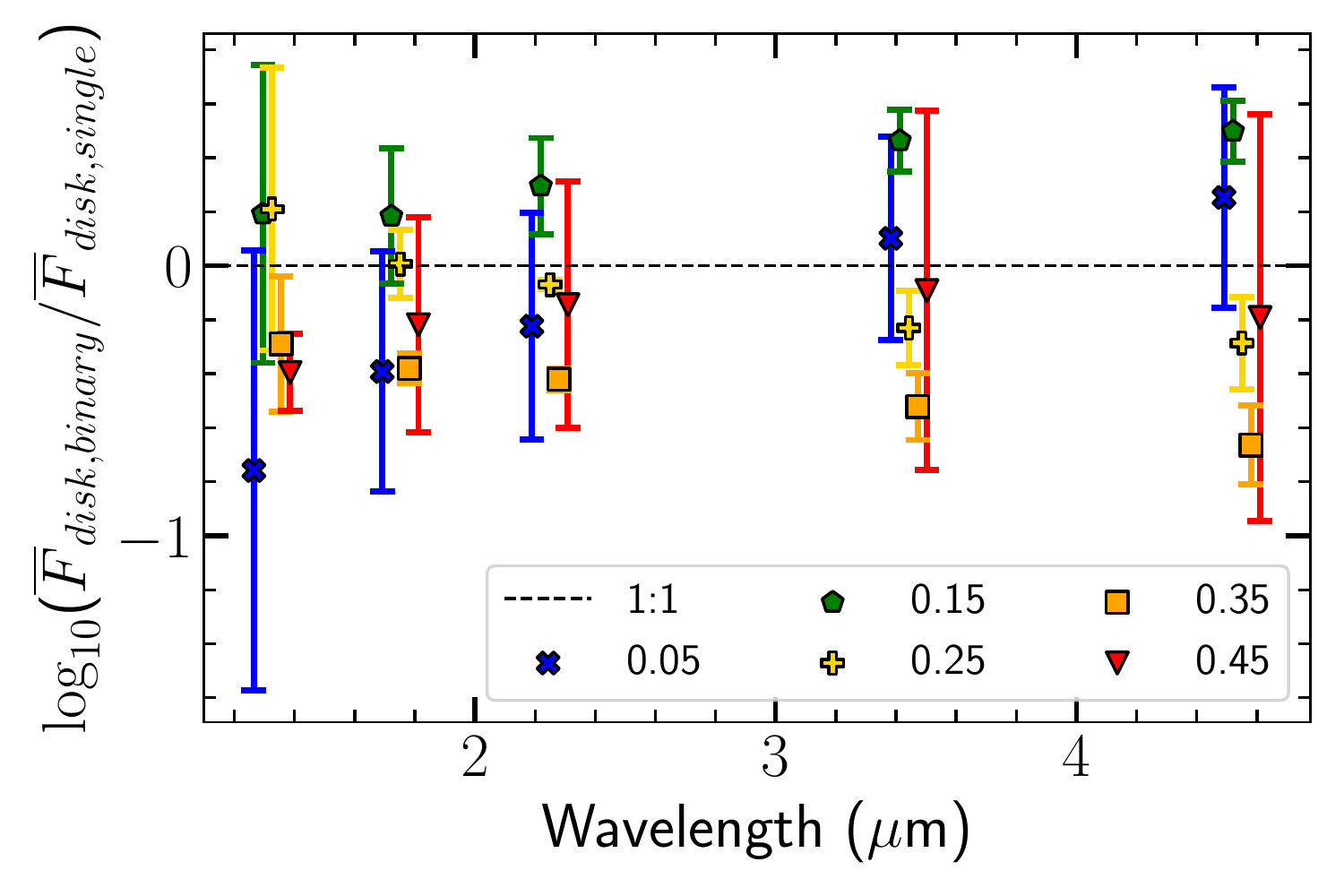}
    \caption{A scatter plot of the average ratio between single and binary star disk fluxes plotted against wavelength and normalized by stellar bolometric flux, color coded by the central veiling value in each $\Delta r_{7510} = 0.1$ bin. The systems have been binned for ease of interpretation. Low-veiling binary disks are systematically brighter than low-veiling single star disks, while high-veiling binary disks are systematically fainter than single star disks, although there is a large amount of RMS scatter in most bins.}
    \label{fig:flux_filter_ratios}
\end{figure}

Referring to Figure \ref{fig:veiling_excess_corr}, it appears that at small veiling values the single and binary star samples both have comparable amounts of excess, but at high veiling values binaries have less excess than single stars. To confirm this, Figure \ref{fig:flux_filter_ratios} shows the average ratio between binary and single star disk fluxes, normalized by stellar bolometric flux, as a function of wavelength, binned by veiling value in increments of 0.1. At low veiling values the binary disk SEDs are typically brighter than the single star disks, but at large veiling values the binary disks are fainter than the single disks. The fainter binary disks at high veiling values drive down the disk excess values, reducing the dynamic range of the relations, causing the shallower slope observed in the optical veiling-NIR excess relations (Figures \ref{fig:veiling_color_corr} and \ref{fig:veiling_excess_corr}). There are very few binary or single star systems at high veiling values ($N \leq 5$ for $r > 0.3$), meaning that it is difficult to draw conclusions from this sample. A larger sample of high-veiling sources would help determine whether this effect is caused by random outliers or physical properties of binaries.

It appears that in our diverse sample of binaries there is little, if any, difference in disk and accretion properties between single and binary stars. The similar distributions of NIR excesses and SED properties between binary and single stars in our sample suggest that the impacts of multiplicity do not create substantial differences between the total populations of binary and single stars. However, many of the effects of multiplicity are only present at small separations (e.g., reduced disk lifetimes only occur at separations $\rho\lesssim 50$ au; \citealt{Cieza2009, Kraus2012a}. Our binary analysis was limited by a small range of veiling values and by a relatively small sample size that did not allow us to investigate the effects of different binary properties on disk attributes. A consistently observed and characterized sample with more systems and a larger range of veiling values is needed to robustly investigate the impacts of multiplicity on inner dust disk properties. For example, similar analyses to HH14 and this work in a larger star-forming region could integrate a much larger binary sample, and such a work could investigate disk properties at different binary separations and mass ratios, which is an important next step in understanding the effect of multiplicity on inner disk properties.

\section{Conclusions}\label{sec:conclusion}
To explore the relationship between NIR excess and optical veiling in T Tauri stars, we have assembled an archival sample of previously-characterized Taurus stars from HH14. We combined the measured properties of the sample with 2MASS and WISE NIR photometry to construct NIR SEDs for our sources and to identify disk hosts, and we used several direct-imaging surveys to determine multiplicity for all the stars in the sample. 

Using this data set, we found that NIR excess (JHK$_{s}$W1W2) is correlated with optical veiling in both colors and magnitudes of excess, suggesting that there is a relationship between the accretion process and the properties of the dust disk's inner wall. The relationship persists in asynchronous measurements, implying that the relationship, if it is causal, is relatively steady-state. One possible explanation is that a more massive inner disk, which could cause a higher inner disk wall, is related to higher accretion rates, causing both increased NIR emission and increased optical veiling. 

We found that the $\sim$50\% of our disk-host sample with no measured veiling showed progressively more asymmetrical NIR excess distributions at redder wavelengths, indicating that they still have circumstellar material (consistent with their disk-host status), but that it may be cooler than the stars with nonzero veiling. This indicates that the inner disk is not as irradiated when there is no veiling, meaning that accretion luminosity may play a significant role in irradiating the inner disk in these sources. Alternatively, some of these objects may be transitional or pre-transitional disks, which have optically thin inner disks while their dust disks remain optically thick at larger radii, and have little or no accretion. Future work investigating the timescale of optical veiling variability and its connection to NIR disk emission is needed to fully understand this effect in the large fraction of apparently veiling-free disk host stars.

We found that binary stars do not show significantly different disk properties than single stars, but note that many of our binary systems were wide binaries, which are known to have similar properties as single stars. Although our high-veiling binary sample suffered from small number statistics, the few high-veiling binaries in our sample had fainter disks than single stars. Future work should focus on a larger sample of close binary stars with a wider range of veiling values to better understand any possible impacts of close companions on inner disk and accretion properties. 

\acknowledgements
We thank Neal Evans, Stella Offner, and Ben Tofflemire for their helpful feedback on the contents of this paper. We thank the referee for their useful comments and suggestions. K.S. acknowledges that this material is based upon work supported by the National Science Foundation Graduate Research Fellowship under Grant No. DGE-1610403. This research has made use of the VizieR catalogue access tool, CDS, Strasbourg, France (DOI : 10.26093/cds/vizier). The original description of the VizieR service was published in 2000, A\&AS 143, 23. This publication makes use of data products from the Wide-field Infrared Survey Explorer, which is a joint project of the University of California, Los Angeles, and the Jet Propulsion Laboratory/California Institute of Technology, funded by the National Aeronautics and Space Administration. This publication makes use of data products from the Two Micron All Sky Survey, which is a joint project of the University of Massachusetts and the Infrared Processing and Analysis Center/California Institute of Technology, funded by the National Aeronautics and Space Administration and the National Science Foundation. This work has made use of data from the European Space Agency (ESA) mission {\it Gaia} (\url{https://www.cosmos.esa.int/gaia}), processed by the {\it Gaia} Data Processing and Analysis Consortium (DPAC, \url{https://www.cosmos.esa.int/web/gaia/dpac/consortium}). Funding for the DPAC has been provided by national institutions, in particular the institutions participating in the {\it Gaia} Multilateral Agreement.

\software{numpy \citep{Harris2020}, matplotlib \citep{Hunter2007}, scipy \citep{Virtanen2020}, astropy \citep{Astropy2013, Astropy2018}, astroquery \citep{Astroquery2019}
}
\bibliography{bib.bib}

\begin{longrotatetable} %
\begin{deluxetable*}{ccCCCccCCCCC}
\setcounter{table}{0}
\tablecaption{System Parameters for Each Source \label{tab:source_params}}
\tablecolumns{11}
\tablewidth{0pt}
\tablehead{
\colhead{Name} & \colhead{2MASS/SIMBAD} & \colhead{Distance} & \colhead{Veiling} & \colhead{\av} & \colhead{Disk} & \colhead{Binary} & \colhead{J} 
& \colhead{H} & \colhead{K$_{s}$} & \colhead{W1} & \colhead{W2}\\
\colhead{} & \colhead{name} & \colhead{(pc)} & \colhead{} & \colhead{(mag)} & \colhead{flag} & \colhead{flag} & \colhead{(mag)} 
& \colhead{(mag)} & \colhead{(mag)} & \colhead{(mag)} & \colhead{(mag)}\\
}
\startdata  
HBC 358 & 2MASS J04034930+2610520 & 148 & 0.00 & 0.05 & N & N & 10.27 & 9.70 & 9.46 & 9.24 & 9.06\\
HBC 359 & 2MASS J04035084+2610531 & 129 & 0.00 & -0.25 & N & N & 10.37 & 9.75 & 9.53 & 9.40 & 9.26\\
HBC 360 & 2MASS J04043936+2158186 & 123 & 0.00 & 0.30 & Y & (4) & 10.80 & 10.17 & 9.97 & 9.84 & 9.69\\
HBC 361 & 2MASS J04043984+2158215 & 124 & 0.00 & 0.40 & Y & (4) & 10.94 & 10.35 & 10.10 & 9.94 & 9.79\\
HBC 362 & 2MASS J04053087+2151106 & 124 & 0.00 & 0.10 & Y & N & 10.95 & 10.29 & 10.06 & 9.89 & 9.78\\
2M 0407+2237 & 2MASS J04073502+2237394 & 126 & 0.00 & 0.80 & Y & N & 12.16 & 11.60 & 11.25 & 11.04 & 10.79\\
LkCa 1 & 2MASS J04131414+2819108 & 127 & 0.00 & 0.45 & N & N & 9.64 & 8.87 & 8.62 & 8.58 & 8.47\\
HBC 366 & 2MASS J04132722+2816247 & 136 & 0.00 & 2.20 & N & N & 8.83 & 7.79 & 7.46 & 7.22 & 7.15\\
V773 Tau & 2MASS J04141291+2812124 & 120 & 0.00 & 0.95 & Y & (2) & 7.49 & 6.64 & 6.21 & 6.33 & 5.81\\
FM Tau & 2MASS J04141358+2812492 & 132 & 0.56 & 0.35 & Y & N & 10.33 & 9.39 & 8.76 & 8.00 & 7.40\\
FN Tau & 2MASS J04141458+2827580 & 130 & 0.03 & 1.15 & Y & N & 9.47 & 8.67 & 8.19 & 7.54 & 6.86\\
CIDA 1 & 2MASS J04141760+2806096 & 135 & 0.13 & 3.00 \tablenotemark{a} & Y & N & 11.73 & 10.58 & 9.88 & 9.13 & 8.26\\
MHO 3 & 2MASS J04143054+2805147 & 180 & 0.00 & 5.30 & Y & (1) & 11.18 & 9.25 & 8.24 & 6.98 & 5.74\\
FP Tau & 2MASS J04144730+2646264 & 127 & 0.06 & 0.60 & Y & (2) & 9.90 & 9.18 & 8.87 & 8.40 & 7.95\\
XEST 20-066 & 2MASS J04144739+2803055 & 129 & 0.00 & -0.15 & Y & N & 10.80 & 10.17 & 9.92 & 9.78 & 9.54\\
CX Tau & 2MASS J04144786+2648110 & 127 & 0.02 & 0.25 & Y & (2) & 9.87 & 9.05 & 8.81 & 8.52 & 8.06\\
LkCa 3 & 2MASS J04144797+2752346 & 131 \tablenotemark{c} & 0.00 & 0.00 & N & (1) & 8.36 & 7.62 & 7.42 & 7.34 & 7.28\\
FO Tau & 2MASS J04144928+2812305 & 136 & 0.07 & 2.05 & Y & (1)(4) & 9.65 & 8.57 & 8.12 & 7.47 & 6.93\\
XEST 20-071 & 2MASS J04145234+2805598 & 134 & 0.00 & 3.00 & N & N & 9.53 & 8.21 & 7.71 & 7.49 & 7.30\\
2M 0415+2818 & 2MASS J04153916+2818586 & 133 & 0.03 & 1.80 & Y & N & 10.55 & 9.61 & 9.23 & 8.75 & 8.29\\
2M 0415+2909 & 2MASS J04154278+2909597 & 160 & 0.00 & 2.80 & Y & N & 10.71 & 9.76 & 9.38 & 9.04 & 8.89\\
2M 0415+2746 & 2MASS J04155799+2746175 & 133 & 0.00 & 0.60 & Y & N & 11.74 & 10.98 & 10.52 & 9.87 & 9.34\\
LkCa 4 & 2MASS J04162810+2807358 & 130 & 0.00 & 0.35 & N & N & 9.25 & 8.52 & 8.32 & 8.25 & 8.13\\
CY Tau & 2MASS J04173372+2820468 & 126 & 0.15 & 0.35 & Y & N & 9.83 & 8.97 & 8.60 & 7.79 & 7.31\\
LkCa 5 & 2MASS J04173893+2833005 & 132 & 0.00 & 0.05 & N & (1) & 9.98 & 9.29 & 9.05 & 8.97 & 8.85\\
V410 X-ray 1 & 2MASS J04174965+2829362 & 131 & 0.04 & 1.70 & Y & N & 11.02 & 9.73 & 9.08 & 8.55 & 7.86\\
V410 X-ray 3 & 2MASS J04180796+2826036 & 125 & 0.00 & 0.20 & Y &  & 11.54 & 10.82 & 10.45 & 10.21 & 9.88\\
V409 Tau & 2MASS J04181078+2519574 & 130 & 0.00 & 1.00 \tablenotemark{a} & Y & N & 10.74 & 9.59 & 9.03 & 8.32 & 7.98\\
HBC 372 & 2MASS J04182147+1658470 & 181 & 0.00 & 0.65 & Y & N & 11.18 & 10.60 & 10.46 & 10.40 & 10.43\\
KPNO 11 & 2MASS J04183030+2743208 & 128 & 0.00 & -0.20 & Y & N & 11.89 & 11.27 & 11.01 & 10.81 & 10.53\\
DD Tau & 2MASS J04183112+2816290 & 127 & 0.44 & 0.75 & Y & (1) & 9.83 & 8.68 & 7.88 & 6.82 & 5.79\\
CZ Tau & 2MASS J04183158+2816585  & 131\tablenotemark{c} & 0.00 & 0.50 & Y  & (1) & 10.52 & 9.77 & 9.36 & 8.69 & 7.50\\
V892 Tau & 2MASS J04184061+2819155 & 134 & 0.00 & 9.30 & Y & N & 8.74 & 7.02 & 5.79 & 5.14 & 4.10\\
Hubble 4 & 2MASS J04184703+2820073 & 127 & 0.00 & 1.35 & N & (1) & 8.56 & 7.64 & 7.29 & 7.13 & 7.02\\
HBC 376 & 2MASS J04185170+1723165 & 122 & 0.00 & 0.25 & N & N & 10.03 & 9.42 & 9.27 & 9.16 & 9.16\\
V410 X-ray 6 & 2MASS J04190110+2819420 & 128 & 0.00 & 1.40 & Y & N & 10.53 & 9.60 & 9.13 & 8.98 & 8.64\\
FQ Tau & 2MASS J04191281+2829330 & 131 \tablenotemark{c} & 0.15 & 1.60 \tablenotemark{a} & Y & (1) & 10.49 & 9.70 & 9.31 & 8.93 & 8.44\\
BP Tau & 2MASS J04191583+2906269 & 127 & 0.30 & 0.45 & Y & N & 9.10 & 8.22 & 7.74 & 7.11 & 6.62\\
V819 Tau & 2MASS J04192625+2826142 & 129 & 0.00 & 1.00 & Y & N & 9.50 & 8.65 & 8.42 & 8.27 & 8.17\\
FR Tau & 2MASS J04193545+2827218 & 126 & 0.03 & 0.20 & Y & N & 10.95 & 10.37 & 9.97 & 9.49 & 8.74\\
LkCa 7 & 2MASS J04194127+2749484 & 125 & 0.00 & 0.05 & N & (1) & 9.13 & 8.38 & 8.26 & 8.15 & 8.03\\
2M 0420+2804 & 2MASS J04202606+2804089 & 128 & 0.00 & 0.25 & Y & N & 10.61 & 9.95 & 9.70 & 9.48 & 9.05\\
XEST 16-045 & 2MASS J04203918+2717317 & 133 & 0.00 & -0.05 & Y & N & 10.50 & 9.86 & 9.56 & 9.50 & 9.30\\
J2 157 & 2MASS J04205273+1746415 & 112 & 0.00 & 0.35 & Y & N & 11.62 & 11.04 & 10.78 & 10.58 & 10.36\\
IRAS 04187+1927 & 2MASS J04214323+1934133 & 147 & 0.08 & 3.10 & Y & N & 10.19 & 8.73 & 8.02 & 7.11 & 6.15\\
DE Tau & 2MASS J04215563+2755060 & 128 & 0.05 & 0.35 & Y & N & 9.18 & 8.27 & 7.80 & 7.08 & 6.54\\
HD 283572 & 2MASS J04215884+2818066 & 127 & 0.00 & 0.50 & N & (2) & 7.41 & 7.01 & 6.87 & 6.76 & 6.77\\
FS Tau & 2MASS J04220217+2657304 & 134 & 0.46 & 2.95 & Y & (1) & 10.70 & 9.24 & 8.18 & 7.26 & 6.50\\
LkCa 21 & 2MASS J04220313+2825389 & 117 & 0.00 & 0.30 & Y & (1)(2) & 9.46 & 8.67 & 8.45 & 8.27 & 8.16\\
XEST 11-078 & 2MASS J04221568+2657060 & 301 & 0.00 & 1.55 & Y & N & 13.81 & 12.62 & 12.03 & 11.51 & 10.49\\
CFHT 21 & 2MASS J04221675+2654570 & 158 & 0.27 & 3.75 & Y & N & 11.58 & 10.04 & 9.01 & 7.73 & 6.93\\
FU Tau & Haro 6-7 & 128 & 0.01 & 1.20 & Y & N & 10.78 & 9.94 & 9.32 & 8.60 & 7.82\\
FT Tau & 2MASS J04233919+2456141 & 130 & 0.27 & 1.30 & Y & N & 10.19 & 9.12 & 8.60 & 7.65 & 7.01\\
IRAS 04216+2603 & 2MASS J04244457+2610141 & 159 & 0.37 & 1.90 & Y & N & 10.80 & 9.75 & 9.05 & 8.02 & 7.32\\
J4423 & 2MASS J04244506+2701447 & 129 & 0.00 & 0.25 & Y & N & 11.34 & 10.71 & 10.46 & 10.33 & 10.12\\
IP Tau & 2MASS J04245708+2711565 & 129 & 0.12 & 0.75 & Y & N & 9.78 & 8.89 & 8.35 & 7.71 & 7.26\\
J4872 A & 2MASS J04251767+2617504 & 131 \tablenotemark{c} & 0.00 & 1.20 & N & N & 9.54 & 8.48 & 8.55 & 7.93 & 7.82\\
KPNO 13 & 2MASS J04265732+2606284 & 130 & 0.00 & 1.80 & Y & N & 11.28 & 10.17 & 9.58 & 8.91 & 8.25\\
DF Tau & 2MASS J04270280+2542223 & 176 & 0.17 & 0.10 & Y & (1)(4) & 8.17 & 7.26 & 6.73 & 6.03 & 5.23\\
DG Tau & 2MASS J04270469+2606163 & 125 & 0.48 & 1.60 \tablenotemark{b} & Y & N & 8.69 & 7.72 & 6.99 & 6.18 & 5.14\\
HBC 388 & 2MASS J04271056+1750425 & 119 & 0.00 & 0.25 & N & N & 8.78 & 8.39 & 8.30 & 8.20 & 8.22\\
J507 & 2MASS J04292071+2633406 & 130 & 0.00 & 0.50 & N & N & 9.82 & 9.09 & 8.79 & 8.65 & 8.48\\
XEST 15-034 & 2MASS J04293623+2634238 & 49 & 0.00 & 0.20 & Y & N & 11.56 & 10.94 & 10.65 & 10.49 & 10.31\\
DH Tau & 2MASS J04294155+2632582 & 133 & 0.40 & 0.65 & Y & (1) & 9.77 & 8.82 & 8.18 & 7.40 & 7.02\\
DI Tau & 2MASS J04294247+2632493 & 138 & 0.00 & 0.70 & Y & (1)(2) & 9.32 & 8.60 & 8.39 & 8.22 & 8.17\\
IQ Tau & 2MASS J04295156+2606448 & 132 & 0.35 & 0.85 \tablenotemark{b} & Y & N & 9.41 & 8.42 & 7.78 & 7.27 & 6.73\\
CFHT 20 & 2MASS J04295950+2433078 & 132 & 0.00 & 2.30 & Y & N & 11.68 & 10.54 & 9.81 & 9.12 & 8.48\\
UX Tau W & HBC 42 & 142 & 0.00 & 0.40 & Y & (1)(4) & 9.87 & 8.95 & 8.92 & 8.98 & 8.77\\
UX Tau E & HBC 43 & 142 & 0.00 & 0.30 & Y & (1)(4) & 8.62 & 7.96 & 7.55 & 6.85 & 6.21\\
FX Tau & 2MASS J04302961+2426450 & 157 & 0.06 & 0.80 & Y & (1) & 9.39 & 8.40 & 7.92 & 7.40 & 6.98\\
ZZ Tau & 2MASS J04305137+2442222 & 134 & 0.02 & 0.55 & Y & (1)(4) & 9.49 & 8.69 & 8.44 & 8.11 & 7.70\\
ZZ Tau IRS & 2MASS J04305171+2441475 & 106 & 0.10 & 1.70 & Y & (1)(4) & 12.84 & 11.44 & 10.31 & 8.56 & 7.08\\
JH 56 & 2MASS J04311444+2710179 & 127 & 0.00 & 0.35 & Y & N & 9.70 & 9.04 & 8.79 & 8.74 & 8.71\\
V927 Tau & 2MASS J04312382+2410529 & 131\tablenotemark{c} & 0.00 & -0.20 & N & (1) & 9.73 & 9.06 & 8.77 & 8.65 & 8.44\\
LkHa 358 & 2MASS J04313613+1813432 & 140 & 0.35 & 2.80 & Y & N & 12.79 & 10.92 & 9.69 & 8.21 & 7.27\\
XZ Tau & 2MASS J04314007+1813571 & 131\tablenotemark{c} & 0.07 & 1.50 & Y & (1) & 9.39 & 8.15 & 7.29 & 6.29 & 4.86\\
HK Tau & 2MASS J04315056+2424180 & 131 & 0.12 & 2.40 & Y & (1) & 10.45 & 9.25 & 8.59 & 7.82 & 7.37\\
J665 & 2MASS J04315844+2543299 & 155 & 0.00 & 0.40 & Y & N & 10.59 & 9.83 & 9.56 & 9.45 & 9.26\\
V1075 Tau & 2MASS J04320926+1757227 & 144 & 0.00 & 0.25 & N & N & 9.70 & 9.06 & 8.85 & 8.75 & 8.72\\
V827 Tau & 2MASS J04321456+1820147 & 164 & 0.00 & 0.05 & N & (1) & 9.16 & 8.49 & 8.23 & 8.15 & 8.06\\
Haro 6-13E & 2MASS J04321540+2428597 & 129 & 0.27 & 2.23 \tablenotemark{a} & Y & N & 11.24 & 9.32 & 8.10 & 6.62 & 5.81\\
Haro 6-13E & 2MASS J04321540+2428597 & 129 & 0.27 & 2.23 \tablenotemark{a} & Y & N & 11.24 & 9.32 & 8.10 & 6.62 & 5.81\\
V826 Tau & 2MASS J04321583+1801387 & 144 & 0.00 & 0.40 & N & N & 9.07 & 8.43 & 8.25 & 8.06 & 8.03\\
MHO 5 & 2MASS J04321606+1812464 & 149 & 0.00 & -0.20 & Y & N & 11.07 & 10.39 & 10.06 & 9.60 & 9.03\\
CFHT 7 & 2MASS J04321786+2422149 & 127 & 0.00 & 0.20 & Y & (1) & 11.54 & 10.79 & 10.38 & 10.09 & 9.77\\
V928 Tau & 2MASS J04321885+2422271 & 140\tablenotemark{c} & 0.00 & 1.95 & N & (1)(4) &  9.54 & 8.43 & 8.11 & 7.91 & 7.83\\
MHO 6 & 2MASS J04322210+1827426 & 146 & 0.01 & -0.15 & Y & N & 11.71 & 11.02 & 10.65 & 10.20 & 9.74\\
MHO 7 & 2MASS J04322627+1827521 & 148 & 0.00 & -0.20 & Y & N & 11.12 & 10.37 & 10.17 & 10.03 & 9.79\\
GG Tau B & CoKu GG Tau c & 148 & 0.00 & 0.00 & Y & (1) & 11.07 & 10.39 & 9.97 & 9.32 & 8.72\\
GG Tau AB & 2MASS J04323034+1731406 & 116 & 0.07 & 1.05 & Y & (1) & 8.67 & 7.82 & 7.36 & 6.36 & 5.81\\
GG Tau AB & 2MASS J04323034+1731406 & 116 & 0.07 & 1.05 & Y & (1) & 8.67 & 7.82 & 7.36 & 6.36 & 5.81\\
FY Tau & 2MASS J04323058+2419572 & 130 & 0.15 & 3.05 & Y & N & 9.98 & 8.68 & 8.05 & 7.32 & 6.77\\
UZ Tau A & 2MASS J04324303+2552311 & 130 & 0.14 & 0.90 & Y & (1)(2) & 9.14 & 8.12 & 7.35 & 6.25 & 5.48\\
HBC 403 & 2MASS J04324373+1802563 & 148 & 0.00 & 0.85 & Y & N & 10.16 & 9.46 & 9.31 & 9.15 & 9.17\\
JH 112A & 2MASS J04324911+2253027 & 140\tablenotemark{c} & 0.00 & 3.15 & Y & (1) & 10.24 & 8.99 & 8.17 & 7.46 & 6.95\\
JH 112B & ** KSA 21B & 164 & 0.00 & 2.95 & Y & (1) & 11.10 & 9.81 & 9.20 & 8.66 & 8.10\\
GH Tau & 2MASS J04330622+2409339 & 140\tablenotemark{c} & 0.00 & 0.40 & Y & (1)(2)(4) &  9.11 & 8.23 & 7.79 & 7.30 & 6.84\\
V807 Tau & 2MASS J04330664+2409549 & 184 & 0.05 & 0.50 & Y & (1)(2) & 8.15 & 7.36 & 6.96 & 6.47 & 6.02\\
V830 Tau & 2MASS J04331003+2433433 & 130 & 0.00 & 0.45 & N & N & 9.32 & 8.61 & 8.42 & 8.41 & 8.38\\
GI Tau & 2MASS J04333405+2421170 & 129 & 0.05 & 2.05 & Y & (1) & 9.34 & 8.42 & 7.89 & 7.09 & 6.35\\
GK Tau A & 2MASS J04333456+2421058 & 129 & 0.08 & 1.35 & Y & (1) & 9.05 & 8.11 & 7.47 & 6.57 & 5.90\\
IS Tau & 2MASS J04333678+2609492 & 146 & 0.02 & 2.55 & Y & (1)(2)(4) & 10.32 & 9.29 & 8.64 & 7.98 & 7.34\\
HN Tau A & 2MASS J04333935+1751523 & 134 & 0.49 & 1.15 & Y & (1)(2) & 10.70 & 9.47 & 8.38 & 7.23 & 6.38\\
2M 0433+2615 & 2MASS J04334465+2615005 & 162 & 0.00 & 3.20 & Y & N & 11.64 & 10.39 & 9.74 & 8.96 & 8.18\\
DM Tau & 2MASS J04334871+1810099 & 144 & 0.13 & 0.10 & Y & N & 10.44 & 9.76 & 9.52 & 9.46 & 9.30\\
CI Tau & 2MASS J04335200+2250301 & 160 & 0.40 & 1.90 & Y & N & 9.48 & 8.43 & 7.79 & 6.76 & 6.03\\
XEST 17-059 & 2MASS J04335252+2256269 & 140\tablenotemark{c} & 0.00 & 1.00 & N & N & 10.24 & 9.47 & 9.11 & 8.93 & 8.65\\
IT Tau A & 2MASS J04335470+2613275 & 160 & 0.00 & 3.10 & Y & (1) & 9.87 & 8.59 & 7.86 & 7.40 & 6.82\\
J2 2041 & 2MASS J04335546+1838390 & 139 & 0.00 & 0.45 & N & (1) & 10.53 & 9.87 & 9.61 & 9.50 & 9.35\\
JH 108 & 2MASS J04341099+2251445 & 162 & 0.00 & 1.75 & Y & N & 10.59 & 9.74 & 9.43 & 9.37 & 9.26\\
HBC 407 & 2MASS J04341803+1830066 & 253 & 0.00 & 0.80 & Y & (1) & 10.58 & 10.08 & 9.90 & 9.81 & 9.83\\
AA Tau & 2MASS J04345542+2428531 & 135 & 0.09 & 0.40 & Y & N & 9.43 & 8.55 & 8.05 & 7.45 & 6.76\\
HO Tau & 2MASS J04352020+2232146 & 164 & 0.20 & 1.00 & Y & N & 11.20 & 10.24 & 9.73 & 8.91 & 8.43\\
FF Tau & 2MASS J04352089+2254242 & 161 & 0.00 & 2.00 & N & (1) & 9.78 & 8.93 & 8.59 & 8.45 & 8.40\\
HBC 412 & 2MASS J04352450+1751429 & 140\tablenotemark{c} & 0.00 & 0.30 & N & (1) & 10.03 & 9.33 & 9.08 & 8.99 & 8.87\\
DN Tau & 2MASS J04352737+2414589 & 129 & 0.00 & 0.55 & Y & N & 9.14 & 8.34 & 8.02 & 7.66 & 7.22\\
HQ Tau & 2MASS J04354733+2250216 & 161 & 0.00 & 2.60 \tablenotemark{a} & Y & N & 8.65 & 7.73 & 7.14 & 6.63 & 5.41\\
HP Tau & 2MASS J04355277+2254231 & 171 & 0.16 & 3.15 & Y & (1) & 9.55 & 8.47 & 7.62 & 6.02 & 5.73\\
HP Tau G2 & 2MASS J04355415+2254134 & 167 & 0.00 & 2.55 & Y & (1) & 8.10 & 7.49 & 7.23 & 7.12 & 7.07\\
Haro 6-28 & 2MASS J04355684+2254360 & 149 & 0.15 & 2.85 \tablenotemark{a} & Y & (1) & 11.14 & 10.06 & 9.53 & 8.82 & 8.20\\
XEST 09-042 & 2MASS J04355892+2238353 & 167 & 0.05 & 1.05 & N & N & 9.32 & 8.60 & 8.37 & 8.20 & 8.17\\
LkCa 14 & 2MASS J04361909+2542589 & 128 & 0.00 & 0.00 & N & N & 9.34 & 8.71 & 8.58 & 8.47 & 8.49\\
2M 0436+2351 & 2MASS J04362151+2351165 & 115 & 0.01 & -0.20 & Y & N & 13.16 & 12.54 & 12.24 & 11.95 & 11.48\\
GM Tau & 2MASS J04382134+2609137 & 139 & 0.26 & 2.10 \tablenotemark{a} & Y & N & 12.80 & 11.59 & 10.63 & 9.54 & 8.85\\
DO Tau & 2MASS J04382858+2610494 & 139 & 0.57 & 0.78 & Y & N & 9.47 & 8.24 & 7.30 & 6.34 & 5.43\\
HV Tau & HBC 418 & 138 & 0.00 & 1.40 & Y & (1)(4) & 9.23 & 8.28 & 7.91 & 7.80 & 7.58\\
2M 0439+2336 & 2MASS J04390163+2336029 & 127 & 0.01 & -0.20 \tablenotemark{a} & Y & N & 11.34 & 10.60 & 10.19 & 9.87 & 9.48\\
VY Tau & 2MASS J04391741+2247533 & 153 & 0.02 & 0.60 & Y & (1) & 9.97 & 9.26 & 8.96 & 8.59 & 8.19\\
LkCa 15 & 2MASS J04391779+2221034 & 157 & 0.12 & 0.30 & Y & N & 9.42 & 8.60 & 8.16 & 7.50 & 7.21\\
GN Tau & 2MASS J04392090+2545021 & 140\tablenotemark{c} & 0.24 & 3.05 & Y & (1) & 10.20 & 8.89 & 8.06 & 7.16 & 6.53\\
ITG 15 & 2MASS J04394488+2601527 & 140 & 0.00 & 2.65 & Y & N & 10.64 & 9.52 & 8.95 & 8.46 & 7.85\\
JH 223 & 2MASS J04404950+2551191 & 142 & 0.00 & 1.20 & Y & (1) & 10.75 & 9.92 & 9.49 & 8.94 & 8.47\\
Haro 6-32 & 2MASS J04410424+2557561 & 140\tablenotemark{c} & 0.00 & 0.75 & Y & N & 10.95 '&10.26 & 9.95 & 9.76 & 9.52\\
IW Tau & 2MASS J04410470+2451062 & 137 & 0.00 & 0.40 & N & (1) & 9.24 & 8.48 & 8.27 & 8.21 & 8.11\\
CoKu Tau 4 & 2MASS J04411681+2840000 & 155 & 0.00 & 1.75 & Y & (1) & 10.16 & 9.08 & 8.66 & 8.50 & 8.35\\
2M 0441+2301 & 2MASS J04414565+2301580 & 124 & 0.00 & -0.15 & Y & N & 10.74 & 10.10 & 9.85 & 9.69 & 9.45\\
HBC 422 & 2MASS J04420548+2522562 & 148 & 0.00 & 2.60 & Y & N & 9.79 & 8.66 & 8.23 & 7.98 & 7.78\\
HBC 423 & 2MASS J04420732+2523032 & 140\tablenotemark{c} & 0.00 & 2.65 & Y & N & 9.58 & 8.40 & 7.95 & 6.91 & 5.38\\
CIDA 7 & 2MASS J04422101+2520343 & 141 & 0.04 & 1.10 & Y & N & 11.40 & 10.57 & 10.17 & 9.68 & 9.05\\
GO Tau & 2MASS J04430309+2520187 & 142 & 0.10 & 1.50 & Y & N & 10.71 & 9.78 & 9.33 & 8.97 & 8.54\\
CIDA 14 & 2MASS J04432023+2940060 & 166 & 0.00 & -0.20 & Y & N & 10.40 & 9.73 & 9.41 & 9.10 & 8.63\\
RX J0446.7+2459 & 2MASS J04464260+2459034 & 148 & 0.00 & 0.00 & Y & (1)(3) & 11.26 & 10.67 & 10.34 & 10.17 & 9.92\\
DQ Tau & 2MASS J04465305+1700001 & 195 & 0.06 & 1.40 & Y & (2) & 9.51 & 8.54 & 7.98 & 7.09 & 6.42\\
DS Tau & HBC 75 & 158 & 0.46 & 0.25 & Y & (2) & 9.47 & 8.60 & 8.04 & 7.35 & 6.84\\
UY Aur & 2MASS J04514737+3047134 & 152 & 0.07 & 1.00 & Y & (1) & 9.13 & 7.99 & 7.24 & 6.11 & 5.04\\
ST 34 & 2MASS J04542368+1709534 & 152 & 0.14 & 0.50 & Y & N & 10.69 & 10.08 & 9.79 & 9.60 & 9.43\\
GM Aur & 2MASS J04551098+3021595 & 158 & 0.17 & 0.30 & Y & N & 9.34 & 8.60 & 8.28 & 8.30 & 8.10\\
LkCa 19 & 2MASS J04553695+3017553 & 157 & 0.00 & 0.50 & N & (2) & 8.87 & 8.32 & 8.15 & 8.06 & 8.06\\
2M 0455+3019 & 2MASS J04554535+3019389 & 157 & 0.01 & 0.70 & Y & N & 11.44 & 10.79 & 10.46 & 10.14 & 9.64\\
AB Aur & 2MASS J04554582+3033043 & 156 & 0.00 & 0.55 & Y & N & 5.94 & 5.06 & 4.23 & 3.29 & 1.67\\
2M 0455+3028 & 2MASS J04554757+3028077 & 147 & 0.00 & 0.20 & Y &  & 11.05 & 10.31 & 9.98 & 9.65 & 9.38\\
XEST 26-062 & 2MASS J04555605+3036209 & 176 & 0.01 & 0.85 & Y & N & 10.47 & 9.66 & 9.27 & 8.85 & 8.37\\
SU Aur & 2MASS J04555938+3034015 & 157 & 0.00 & 0.65 & Y & N & 7.20 & 6.56 & 5.99 & 5.18 & 4.47\\
HBC 427 & 2MASS J04560201+3021037 & 165 & 0.00 & 0.20 & Y & (1) & 8.96 & 8.32 & 8.13 & 7.89 & 7.86\\
V836 Tau & 2MASS J05030659+2523197 & 167 & 0.02 & 0.60 & Y & (2) & 9.91 & 9.08 & 8.60 & 8.19 & 7.72\\
CIDA 8 & 2MASS J05044139+2509544 & 169 & 0.10 & 1.70 & Y & N & 10.91 & 10.01 & 9.60 & 9.22 & 8.79\\
CIDA 9 & 2MASS J05052286+2531312 & 175 & 0.10 & 0.70 \tablenotemark{a} & Y & (1) & 12.81 & 11.91 & 11.16 & 9.17 & 8.29\\
CIDA 9 & 2MASS J05052286+2531312 & 175 & 0.10 & 0.70 \tablenotemark{a} & Y & (1) & 12.81 & 11.91 & 11.16 & 9.17 & 8.29\\
CIDA 10 & 2MASS J05061674+2446102 & 176 & 0.00 & 0.55 & Y & (1) & 10.79 & 10.09 & 9.81 & 9.68 & 9.49\\
CIDA 11 & 2MASS J05062332+2432199 & 167 & 0.05 & 0.35 & Y & (1) & 10.42 & 9.71 & 9.46 & 9.21 & 8.81\\
2M 0506+2104 & 2MASS J05064662+2104296 & 160 & 0.00 & -0.20 & Y & N & 12.05 & 11.41 & 11.11 & 10.90 & 10.65\\
RW Aur A & 2MASS J05074953+3024050 & 183 & 0.52 & -0.25 & Y & (1) & 8.38 & 7.62 & 7.02 & 6.25 & 5.23\\
CIDA 12 & 2MASS J05075496+2500156 & 170 & 0.03 & 0.50 & Y & N & 11.41 & 10.70 & 10.40 & 10.18 & 9.79\\
\enddata
\tablecomments{Binary flag (1) corresponds to \citet{Kraus2011}, (2) is \citet{Daemgen2015}, (3) is \citet{Kraus2012b}, and (4) is \citet{Schaefer2014}. The SIMBAD name is the searchable name on the SIMBAD online service. The veiling and extinction are the values from HH14.}
\tablenotetext{a}{\av\ measurement from HH14 may be incorrect or require R$_{V} \neq 3.1$, and system was removed from analysis.}
\tablenotetext{b}{System has an anomalous SED and was removed from analysis.}
\tablenotetext{c}{System did not have a \textit{Gaia} distance measurement, so the HH14 distance measurement was used.}
\end{deluxetable*}
\end{longrotatetable}

\end{document}